\documentclass[a4paper,11pt]{article}
\usepackage{jcappub} % for details on the use of the package, please see the JINST-author-manual
\usepackage{multirow}
\usepackage[utf8]{inputenc}
\usepackage[T1]{fontenc}
\usepackage{subcaption}
\usepackage{caption}
\usepackage{booktabs}

\usepackage[top=2.5cm,bottom=3.cm,right=2.5cm,left=2.5cm]{geometry}
\usepackage{url}

\usepackage{soul,xcolor}
\setstcolor{red}

\newcommand{\diff}{\mathrm{d}}

\title{\fontsize{20}{32}\selectfont{Testing parity with composite-field spectra of BOSS and DESI luminous red galaxies }} 

\subheader{
\footnotesize
\vspace*{-3.7em}
\begin{flushright}
LAPTH-020/26 \\
RBI-ThPhys-2025-nn
\end{flushright}
}

\author[a]{\fontsize{13.84}{25}\selectfont Zucheng Gao,}
\author[a]{Marina S. Cagliari,}
\author[a]{Azadeh Moradinezhad Dizgah,}
\author[b]{Zvonimir Vlah}

\affiliation[a]{\vspace{0.3cm} Laboratoire d’Annecy de Physique Theorique (LAPTh), CNRS/USMB, 9 Chemin de Bellevue BP110 - Annecy -  F-74941 - ANNECY CEDEX - FRANCE}
\affiliation[b]{Division of Theoretical Physics, 
Ru\dj er Bo\v{s}kovi\'{c} Institute, Zagreb HR-10000, Croatia}
\emailAdd{zucheng.gao@lapth.cnrs.fr, marina.cagliari@lapth.cnrs.fr, azadeh.moradinezhad@lapth.cnrs.fr, zvlah@irb.hr}

\abstract{Detection of parity violation on cosmological scales would have profound implications for fundamental physics. Motivated in part by recent measurements of parity-odd four-point correlation functions in BOSS and DESI luminous red galaxy samples, which probe parity violation in the scalar sector, we present the first measurement of parity-odd kurto spectra in spectroscopic galaxy survey data. We analyse two composite-field spectra, $\mathcal{P}_{2\times2}$ (vector--pseudo-vector) and $\mathcal{P}_{3\times1}$ (scalar--pseudo-scalar). Compared with parity-odd four-point correlation function analyses, the kurto-spectrum formalism performs physically motivated compression on the trispectrum into a substantially lower-dimensional data vector, allowing direct estimation of covariance matrices from mock catalogues and reducing sensitivity to covariance-modelling systematics. Using null-hypothesis $\chi^2$ tests and cross-patch consistency checks, we find no evidence for a cosmological parity-violating signal in either survey. We examine the impact of the adopted mock catalogues and find that the high-fidelity mocks provide a better match to the data of both surveys than the approximate mocks. The DESI DR1 measurements exhibit a scatter smaller than that of BOSS DR12 by about a factor of four, consistent with the improved statistical precision expected from the higher tracer number density. Future DESI data releases, with larger volume and number density, together with larger suites of high-fidelity mocks, can enable significantly sharper tests of parity violation using kurto spectra.}

\begin{document}

\date{\today}
\maketitle

\vspace{0.1in}

\section{Introduction}\label{sec:intro}

A detection of a parity-odd cosmological signal would point either to new fundamental physics operating in the early or late Universe, or to astrophysical sources. In the scalar sector, primordial parity violation can arise in a variety of inflationary scenarios, including ghost inflation and related non-Bunch--Davies setups with higher-derivative interactions~\cite{Arkani-Hamed:2003juy,Cabass:2022oap,Cabass:2022rhr}, cosmological-collider models with massive spinning fields~\cite{Arkani-Hamed:2015bza,Liu:2019fag,Cabass:2022rhr,Cabass:2022oap,Jazayeri:2023kji}, and chiral gauge-field models during inflation~\cite{Bartolo:2015dga,Niu:2022fki,Fujita:2023inz,Stefanyszyn:2023qov,Shiraishi:2016mok,Cho:2025rvg}. Parity-odd scalar fluctuations may also be generated by helical primordial magnetic fields~\cite{Shaw:2009nf,Caprini:2009pr,Caprini:2014mja,Caprini:2017vnn,Bamba:2021wyx,Yura:2025mus}. Parity violation may also arise in vector and tensor sectors~\cite{Zhu:2024wme,Shim:2024tue} through mechanisms like chiral gravity or chiral gauge dynamics~\cite{Lue:1998mq,Jackiw:2003pm,Alexander:2007kv,Anber:2012du,Adshead:2013qp}, with observable signatures including parity-odd cosmic microwave background (CMB) polarization~\cite{Gerbino:2016mqb}, cosmic birefringence~\cite{Minami:2020odp,Fujita:2020ecn,Cosmoglobe:2023pgf}, cosmic shear~\cite{Biagetti:2020lpx,Philcox:2023uor,Kurita:2025hmp}, and tensor or mixed scalar--tensor higher-point functions~\cite{Bartolo:2017szm,Bartolo:2018elp,Philcox:2023xxk}. Tensor-sector parity violation can also induce scalar-sector imprints~\cite{Masui:2010cz,Jeong:2012df,Schmidt:2013gwa,Masui:2017fzw,Creque-Sarbinowski:2023wmb,Inomata:2024ald}. 
In this work, we focus on testing parity violation in the scalar sector through observations of large-scale structure.

In the scalar sector, parity violation first appears at leading order in the four-point correlation function (4PCF) and its Fourier-space counterpart, the trispectrum. Lower-order statistics cannot encode a scalar parity-violating signal, since the corresponding parity transformation can be reproduced by a rotation in three-dimensional space and is therefore preserved by statistical isotropy. A parity-violating scalar trispectrum is imaginary and contains an odd number of Levi-Civita tensors. Both statistics are high-dimensional, making their measurement and covariance estimation difficult, which are the key challenges in uncovering parity-violation features.

In the context of galaxy surveys, several recent studies have attempted to probe parity-violating (PV) signatures using the parity-odd 4PCF. Using the Baryon Oscillation Spectroscopic Survey Data Release 12 (BOSS-DR12) luminous red galaxy (LRG) samples, \cite{Philcox:2022PRD} reported a $2.9\sigma$ detection based on a non-parametric rank test comparing the data pseudo-$\chi^2$ to the distribution obtained from parity-conserving mocks, with the pseudo-$\chi^2$ weighted by a fiducial analytic covariance. The same work also performed a compressed analysis, in which the analytic covariance was used to define a reduced basis and the covariance of the compressed statistic was then estimated from mocks. An independent analysis by \cite{Hou:2023MNRAS} found detections at $3.1\sigma$ in the LOWZ sample and $7.1\sigma$ in the CMASS (constant stellar mass) sample using three complementary approaches: a direct analysis with an analytic covariance calibrated against mocks, a compressed analysis with mock-estimated covariance in the reduced basis, and a reduced-$\ell_{\max}$ analysis in which the dimensionality is lowered sufficiently to allow a direct mock-covariance treatment. A later reanalysis of BOSS-DR12 using improved mocks reduced the inferred significance to $1.4\sigma$~\cite{Philcox:2024mmz}.

These results were revisited by~\cite{Krolewski:2024JCAP}, who showed that the apparent detections can arise from a data--mock mismatch that biases the standard $\chi^2$ statistic used in previous analyses when the parity-even 8PCF of the data differs from that of the mocks. To separate a genuine parity-violating signal from this bias term, they introduced two new statistics: a signal-sensitive cross-patch statistic, $\chi^2_{\times}$, and a null statistic, $\chi^2_{\Delta}$, which isolates the stochastic mismatch contribution. Applying these statistics to BOSS-DR12, they found that the inferred PV significance ranges from $0$ to $2.5\sigma$, while the 8PCF bias term is detected at $\sim 6\sigma$, and concluded that there is no compelling evidence for parity violation in the BOSS data.

More recently, analyses of the Dark Energy Spectroscopic Instrument (DESI) DR1 have reported apparent $4$--$10\sigma$ detections in the parity-odd 4PCF measured within individual survey patches~\cite{Slepian:2025kbb,Hou:2025cey}. The two analyses differ in their treatment of the covariance and in the choice of null statistics, but both emphasise that the interpretation of the apparent excess is sensitive to these choices. In particular, cross-patch analyses do not support a statistically significant shared PV signal across the sky, and accounting for data--mock covariance mismatch reduces the significance of the excess. Taken together, the current DESI DR1 results do not provide robust evidence for parity violation.

The central challenge in the 4PCF approach lies in the estimation of the covariance matrix. Because the dimensionality of the 4PCF is much larger than the number of available mocks, direct numerical estimation of the full covariance is generally not robust. In many of these analyses, the covariance treatment therefore relied at least partly on analytic covariance models, while mocks were used to calibrate, validate, or estimate the covariance of compressed statistics. However, such analytic covariances rely on Gaussian-field assumptions and do not fully capture the non-Gaussian and nonlinear contributions to the true covariance, which can lead to a misestimation of the statistical significance.

A powerful way to address these challenges is through the composite-field (CF) spectra framework, which compresses the information contained in $N$-point correlation functions into two-point statistics. Constructed from cross-correlations of appropriately weighted products of the observed overdensity field, CF spectra are designed to isolate specific physical signatures in polyspectra. Their much lower dimensionality relative to the full $N$-point function makes them computationally efficient and allows more robust covariance estimation from simulations. At lowest order, for example, a set of galaxy skew spectra capture the full information of bispectrum on large scales, providing the same constraints on galaxy bias parameters, the growth rate of structure, and primordial non-Gaussianity~\cite{Schmittfull:2014tca,MoradinezhadDizgah:2019xun,Schmittfull:2020hoi}. For other cosmological parameters, including $\Lambda$CDM parameters and neutrino mass, the constraints from skew spectra have been shown to be competitive with those from the bispectrum in both simulation-based forecasts~\cite{Hou:2022rcd} and analyses of BOSS data~\cite{Hou:2022rcd}. Extending this framework to the trispectrum level, kurto spectra were originally introduced to probe the third-order halo biases generated by nonlinear gravitational evolution~\cite{Lazeyras:2017hxw,Abidi:2018eyd}.

The same framework was recently extended to probe PV trispectra~[\citealp{Gao:2025yqd}; see also \citealp{Jamieson:2024mau}]. By combining the observed scalar overdensity field into vector and pseudo-vector fields, or scalar and pseudo-scalar fields, one can construct cross-power spectra whose expectation values are non-zero only in the presence of a parity-violating signal, while cancelling the contribution from parity-even trispectrum components. In our recent work~\cite{Gao:2025yqd}, we extensively validated the properties of these parity-odd CF estimators using both analytic calculations and numerical simulations, showing that the dominant source of noise arises from parity-even stochastic contributions and decreases with increasing tracer number density. Given that the LRG samples in BOSS-DR12 and DESI-DR1 have a higher mean number density than the \textsc{Quijote-ODD} simulations used in~\cite{Gao:2025yqd}, we expect tighter constraints on the amplitude of any parity-violating signal.

In this work, we measure the parity-odd kurto spectra in both BOSS-DR12 and DESI-DR1, together with the corresponding mock catalogues, providing an alternative and potentially more robust approach to searching for parity-violating signatures than previous 4PCF analyses. The covariance matrices are estimated numerically from the mocks, and we perform null-hypothesis tests by comparing the data $\chi^2$ values to the corresponding mock $\chi^2$ distributions. We further investigate the impact of different mock suites on the covariance estimation. In addition, cross-patch analyses are performed in both datasets to test the consistency of any apparent signal and to assess whether the data are compatible with the null hypothesis.

The rest of the paper is organised as follows: in section~\ref{sec:Methods and data}, we first describe the mathematical forms of the two kurto spectra and the overdensity field definition from data and mocks. In section~\ref{subsec:boss_data} and~\ref{subsec:desi_data}, we review the BOSS-DR12 and DESI-DR1 data, along with their corresponding mocks. We define the null-hypothesis test in section~\ref{sec:chi2}. In section~\ref{sec:results}, we present our results and analysis. And we conclude in section~\ref{sec:conclusion}.

\section{Methods and Data} \label{sec:Methods and data}

In this section, we present the kurto spectrum estimator, define the null-hypothesis test used to assess the presence of the PV signal, and describe the datasets and mock catalogues employed in the analysis.

\subsection{General construction of parity-odd kurto spectra}

Composite-fields in configuration space, denoted as ${\mathcal D}_n[\delta](\mathbf{x})$, are defined as products of filtered density fields evaluated at the same position,
\begin{equation}\label{eq:CF_def_config}
{\mathcal D}_n[\delta](\mathbf{x}) = \prod_{i=1}^{n} {\mathcal D}_i(\mathbf{x})\delta(\mathbf{x}),
\end{equation}
where ${\mathcal D}_i$ are arbitrary operators. The Fourier transformed composite-field of order $n$ is given by a convolution:
\begin{align}\label{eq:CF_def_Fourier}
    D_n[\delta](\mathbf{k}) &= \int_{\mathbf{q}_1} \cdots \int_{\mathbf{q}_{n-1}} D_n(\mathbf{q}_1,\cdots,\mathbf{q}_{n-1},\mathbf{k}-\mathbf{q}_1-\cdots-\mathbf{q}_{n-1})  \notag \\ & \hspace{0.8in}  \times \delta(\mathbf{q}_1)  \cdots\delta(\mathbf{q}_{n-1})\delta(\mathbf{k}-\mathbf{q}_1-\cdots-\mathbf{q}_{n-1}), 
\end{align}
where $D_n(\mathbf{q}_1,\cdots,\mathbf{q}_{n-1},\mathbf{k}-\mathbf{q}_1-\cdots-\mathbf{q}_{n-1})$ is the coupling kernel for $n$ fields in Fourier space. Since the convolution in eq. \eqref{eq:CF_def_Fourier} receives contributions from small-scale fluctuations, which can be affected by modelling uncertainties or systematics,  we smooth the density field entering the Fourier-space composite fields, as described below.

The CF two-point correlator is then defined as
\begin{equation}\label{eq:CF_spec}
\langle D_n[\delta](\mathbf{k}) D_m[\delta](\mathbf{k}')\rangle = (2\pi)^3 \delta_D(\mathbf{k}+\mathbf{k}') \mathcal{P}_{D_n[\delta],D_m[\delta]}(\mathbf{k}),
\end{equation}
where $\delta_D(\mathbf{k}+\mathbf{k}')$ is the Dirac-delta function, and $\mathcal{P}_{D_n[\delta],D_m[\delta]}$ is the CF spectrum, 
which encodes the information of $(n+m)$-point statistics. In principle, the $D_n$ and $D_m$ are chosen to match the theoretical template of the target $(n+m)$-point function. We point the interested readers to refs.~\cite{Schmittfull:2014tca,Schmittfull:2020hoi,Gao:2025yqd} for a more detailed discussion.

To capture the information of the galaxy trispectrum, two types of kurto spectra can be constructed: correlation of the observed galaxy overdensity field with a cubic composite field or correlations of two quadratic composite fields. To search for possible imprints of parity violation, the kernels $D_n$ can be chosen such that they null the parity-even trispectrum and pick out parity-odd contributions. In this work, we use the kernels derived from a parity-odd trispectrum template considered in~\cite{Coulton:2024}. We thus consider two types of parity-odd kurto spectra as $\mathcal{P}_{2\times 2}^{\rm PO}$ and $\mathcal{P}_{3\times 1}^{\rm PO}$ and report their general form below but refer the interested reader to ref.~\cite{Gao:2025yqd} for further details of construction of parity-odd kurto spectra and more generally composite-field spectra, and here just review the main equations defining the two observables. 

Starting from the observed galaxy overdensity field, $\delta(\mathbf{x})$, the first type of kurto spectra is defined as,
\begin{equation}\label{eq:P_2x2}
\mathcal{P}_{2\times 2}^{\rm PO}(k) = \int \frac{\diff\Omega_k}{4\pi}\langle {\bf V}^{ab}({\bf k}) \cdot {\bf A}^{cd}(-{\bf k}) \rangle,
\end{equation}
where the vector field ${\bf V}^{ab}$ and the pseudo-vector ${\bf A}^{cd}$ are given by
\begin{align}
V_i^{ab}(\mathbf{x}) &= \delta^a(\mathbf{x}) \, \partial_{i}\delta^b(\mathbf{x}) \, , \label{eq:V_conf}\\ 
A_i^{cd}(\mathbf{x})  &= \epsilon_{ijk} \, \partial_j \partial^{-2} V_k^{cd}(\mathbf{x}) \,, \label{eq:A_conf} 
\end{align}
with the superscription $\{a,b,c,d\}$ indicating different filtering of the field in Fourier-space, which explicitly reads
\begin{align}
\label{eq:VA_Fourier}
V^{ab}_i(\mathbf{k})&= -\mathrm{i} \, \int_{\mathbf q} (\mathbf{k}-\mathbf q)_i \, f_a(q) \, f_b(|\mathbf k-\mathbf q|) \, \delta(\mathbf q) \, \delta(\mathbf k-\mathbf q) \, , \\
A^{cd}_i(\mathbf{k})&= - \frac{\mathrm{i}}{k^{2}} \, \left[\mathbf k \times {\bf V}^{cd}(\mathbf{k})\right]_i \,,
\end{align}
where the $f_{i}$ are the filtering functions(see eq.~\ref{eq:filtering-functions}). We have defined a simplified notation
\begin{equation}
    \int_{\mathbf{q}} \equiv \int \frac{d^3q}{(2\pi)^3}.
\end{equation}

The second kurto spectrum is defined by correlating a pseudo-scalar cubic field (thus sensitive to the parity-odd signature) and the original field (or its smoothed version),
\begin{equation}\label{eq:P_3x1}
\mathcal{P}^{\rm PO}_{3\times1}(k) = \int \frac{\diff\Omega_k}{4\pi} \langle \delta(\mathbf{k}) \, \Psi^{abc}(-\bf k)\rangle \, ,
\end{equation}
where the pseudo-scalar cubic field is constructed as~\cite{Jamieson:2024mau},
\begin{equation}\label{eq:cubic_psudoscalar}
\Psi^{abc}(\mathbf{x}) = \nabla \delta^a(\mathbf{x}) \cdot \mathbf{B}^{bc}(\mathbf{x}) 
\end{equation}
where $\mathbf{B}^{bc}$ is a quadratic pseudo-vector field, 
\begin{equation}
\mathbf{B}^{bc}(\mathbf{x}) = \nabla \delta^b(\mathbf{x})\times \nabla \delta^c(\mathbf{x}) 
\label{eq:B_conf}
\end{equation}
with their Fourier transforms given by\footnote{Ref.~\cite{Jamieson:2024mau} allowed for an additional overall filter for $\bf B$, which in their numerical evaluation of the estimator was set to unity. We therefore do not write this extra filter in the expression below.}
\begin{align}
\Psi^{abc}(\mathbf{k}) &= -\mathrm{i} \int_{\mathbf{q}_1} (\mathbf{k}-\mathbf{q}_1)_i \, f_a(|\mathbf{k}-\mathbf{q}_1|) \, \delta(\mathbf{k}-\mathbf{q}_1) \, B^{bc}_i(\mathbf{q}_1) \, ,\\
B^{bc}_i(\mathbf{q}_1) &= - \epsilon_{ijk} \, \int_{\mathbf{q}_2} \mathbf{q}_{2,j} \, (\mathbf{q}_1-\mathbf{q}_2)_k \, f_b(q_2) \, f_c(|\mathbf{q}_1-\mathbf{q}_2|) \, \delta(\mathbf{q}_2) \, \delta(\mathbf{q}_1-\mathbf{q}_2) \, .
\end{align}
In what follows, we will use the same scalar kernel functions as ref.~\cite{Jamieson:2024mau}, 
\begin{equation}
f_a(k) = k^2, \qquad f_b(k) = f_d(k) = 1, \qquad f_c(k) = k^{-2} \,.
\label{eq:filtering-functions}
\end{equation}
We formally drop the superscription $\mathrm{PO}$ in the rest of this paper for simplicity.

In practice, we smooth the overdensity field entering the Fourier-space composite fields: 
\begin{equation}\label{eq:tanh_filter_field}
    \delta_{W}(\mathbf{x}) = \int \frac{\diff^3 k}{(2\pi)^3} e^{-i \, \mathbf{k} \cdot \mathbf{x}} \, W(k) \, \delta(\mathbf{k}) \, ,
\end{equation}
where we chose Heaviside-step-function as the smoothing window function:
\begin{equation}
    W(k) = \frac{1}{2} \, \left[1 - \tanh\left( \alpha \ln\frac{k}{k_{\rm max}} \right) \right] \, , 
\end{equation}
with $\alpha=50$ and $k_{\rm max}=0.45\, h \, \mathrm{Mpc}^{-1}$, which filters out the comoving scales smaller than $R\sim 7 \, h^{-1} \, \mathrm{Mpc}$. 

\subsection{FKP estimator for parity-odd kurto spectra}

In order to analyse data from spectroscopic galaxy surveys, namely BOSS and DESI, and in analogy with the skew-spectrum analysis of ref.~\cite{Hou:2024blc}, we extend the estimators developed in ref.~\cite{Gao:2025yqd}, which measure the kurto spectra in periodic box simulations using Fast Fourier Transforms, to a formulation suitable for survey data that accounts for survey geometry and observational systematics. 

To this end, we construct an equivalent of Feldman-Kaiser-Peacock \citep[FKP;][]{FKP} estimators for the power spectrum, bispectrum, and skew spectra, by introducing a weighted fluctuation field defined from the observed galaxy density and a random catalogue that traces the survey selection function,
\begin{equation}
    F(\mathbf{x}) = \frac{w_{\mathrm{FKP}}(\mathbf{x})}{I_{44}^{1/4}} 
    \left[ w_{\mathrm{g}}(\mathbf{x}) \, n_{\mathrm{g}}(\mathbf{x}) 
    - \alpha_{\mathrm{r}} \, w_{\mathrm{r}}(\mathbf{x}) \, n_{\mathrm{r}}(\mathbf{x}) \right] \,.
    \label{eq:fluctuation}
\end{equation}
Here, $n_{\mathrm{g}}(\mathbf{x})$ and $n_{\mathrm{r}}(\mathbf{x})$ denote the number densities of galaxies and randoms, while $w_{\mathrm{g}}(\mathbf{x})$ and $w_{\mathrm{r}}(\mathbf{x})$ are the corresponding systematic weights. The function
\begin{equation}
    w_{\mathrm{FKP}}(\mathbf{x}) = \frac{1}{1 + \bar{n}(z) \, P_0}
    \label{eq:FKP}
\end{equation}
is the FKP weight, where $\bar{n}(z)$ is the mean galaxy density at redshift $z$ and $P_0 = 1 \times 10^{4}\, h^{-3} \, \mathrm{Mpc}^{3}$ denotes the characteristic power at the scales of interest, chosen to reduce the variance of the estimator.\footnote{We adopted the FKP weights, which minimise the variance for the power spectrum. This may not be optimal for kurto spectra.} The coefficient
\begin{equation}
    \alpha_{\mathrm{r}} = \frac{\sum_{i=1}^{N_\mathrm{g}} w_{\mathrm{g}, i}}
    {\sum_{i=1}^{N_\mathrm{r}} w_{\mathrm{r}, i}}
    \label{eq:alpha}
\end{equation}
ensures the proper normalisation of the random number density, where the sums are performed object-wise over the galaxies and randoms in the catalogues, with $N_{\mathrm{g}}$ and $N_{\mathrm{r}}$ denoting their total numbers. The systematic weighting schemes adopted for each survey and catalogue are described in sections~\ref{subsec:boss_data} and \ref{subsec:desi_data}. Finally, $I_{44}$ denotes the field normalisation,
\begin{equation}
    I_{44} = \int \diff \mathbf{x} \, 
    w_{\mathrm{FKP}}^4(\mathbf{x}) 
    \left[ w_{\mathrm{g}}(\mathbf{x}) \, n_{\mathrm{g}}(\mathbf{x}) \right]^4 \, ,
    \label{eq:I44}
\end{equation}
for which we discuss different practical approximations in the following two subsections.

We implemented the kurto spectra estimators within the \texttt{nbodykit} framework \citep{2018AJ....156..160H}. Operatively, we start in configuration space by painting the fluctuation field (see eq.~\ref{eq:fluctuation}) on a mesh. Then, we use this mesh to build the composite fields from eqs.~\eqref{eq:V_conf}, \eqref{eq:A_conf}, \eqref{eq:cubic_psudoscalar}, and \eqref{eq:B_conf}. Finally, we utilise the standard power spectrum estimator to measure the two kurto spectra of interest, where we modified the normalisation factor to eq.~\eqref{eq:I44}. This procedure follows the steps of the codes that measure skew spectra \cite{Schmittfull:2020hoi,Hou:2024blc}.

\subsection{BOSS-DR12 data and mocks}\label{subsec:boss_data}

\begin{figure}
    \centering
    \includegraphics[width=0.65\linewidth]{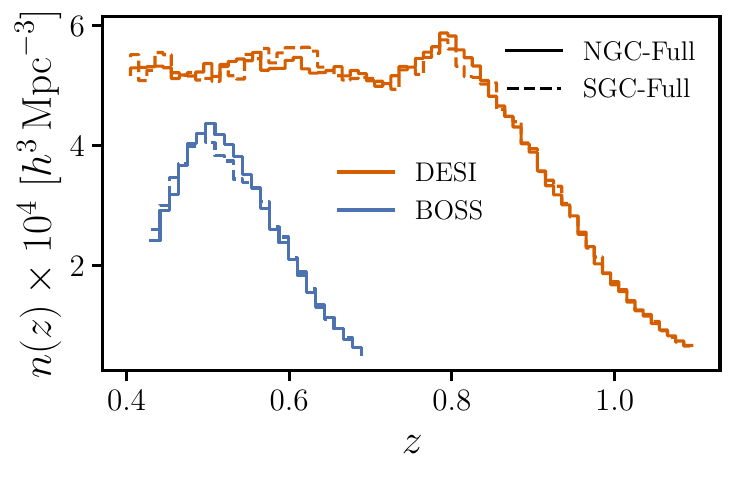}
    \caption{Number density as a function of redshift for BOSS DR12 (blue) and DESI (orange) in the NGC (solid lines) and SGC (dashed lines).}
    \label{fig:nz_comparison}
\end{figure}

To compare with previous analyses of 4PCF works~\cite{Hou:2023MNRAS, Philcox:2022PRD, Krolewski:2024JCAP}, we start from the kurto spectra analysis of the BOSS-DR12 LRGs. 

The survey is divided into two fields of view, the North Galactic Cap (NGC) with a sky area of $7429 \, \mathrm{deg}^2$ and the South Galactic Cap (SGC) with an area of $2823 \, \mathrm{deg}^2$ \cite{BOSS-dr12-lss}. For the analysis, we use the combination of the CMASS and the LOWZTOT catalogues in the redshift range $0.43 < z < 0.7$. These catalogues contain about $800\,000$ and $400\,000$ galaxies in the NGC and SGC, respectively. The mean number density is about $3.5 \times 10^{-4} \, h^{3} \, \text{Mpc}^{-3}$, but it shows large variations within the redshift range of interest, as shown in figure~\ref{fig:nz_comparison}. To convert the observed angles and redshifts to distances in BOSS-DR12 measurements, we assume the fiducial cosmology of $\Omega_{\mathrm{m}} = 0.31$, $h=0.676$, $\Omega_{\mathrm{b}} h^2 = 0.022$, $\sigma_8=0.8$, and $n_{\mathrm{s}} = 0.97$ \cite{Alam:2017MNRAS}.

For each field of view, we use the random catalogues provided by the collaborations, which are 50 times denser than the corresponding galaxy catalogues \cite{BOSS-dr12-lss}. The observed galaxies and randoms are assigned systematic weights used to define the fluctuation field. More specifically, the galaxy catalogue has the following weighting scheme
\begin{equation}
    w_{\mathrm{g}}^{\mathrm{DR12}} = w_{\rm sys} \, (w_{\rm noz} + w_{\rm cp}-1) \, ,
    \label{eq:weight_g-dr12}
\end{equation}
where $w_{\rm noz}$ and $w_{\rm cp}$ are the redshift-failure and fiber-collision weights,  and $w_{\rm sys}$ is the imaging systematic weights. 
The random catalogue does not have systematics weights, hence
\begin{equation}
    w_{\mathrm{r}}^{\mathrm{DR12}} = 1 \, .
    \label{eq:weight_r-dr12}
\end{equation}

For the data covariance estimation and the null-hypothesis test, we use two sets of mock catalogues: the \textsc{MultiDark-Patchy} mocks~\citep[hereafter \textsc{Patchy};][]{Kitaura:2015uqa}, which are the official BOSS-DR12 mocks, and the \textsc{GLAM-Uchuu} mocks~\citep[hereafter \textsc{Uchuu};][]{Ereza:2023zmz}. 
The two suites differ in the modelling of the underlying matter field, in the prescription used to generate the galaxy distribution, and in the construction of the lightcones. Using both suites allows us to test the sensitivity of the null-hypothesis analysis to the adopted mock catalogue. This is similar in spirit to the BOSS 4PCF study of ref.~\cite{Philcox:2024mmz}; however, unlike the 4PCF case, the much lower dimensionality of the kurto spectra allows us to estimate the covariance directly from the mocks, without relying on an analytic covariance model or additional data compression.

The \textsc{Patchy} mocks combine an approximate treatment of dark matter gravitational evolution with a phenomenological, non-perturbative stochastic biasing model. Their calibration relies on a reference BOSS catalogue built from the \textsc{BigMultiDark} $N$-body simulation through halo abundance matching to the BOSS-DR12 clustering data. Additional redshift-space distortion parameters are then tuned to reproduce the damping of the monopole and quadrupole, first in the \textsc{BigMultiDark} simulation and then in the BOSS-DR12 observations. By contrast, the \textsc{Uchuu} mocks are based on $N$-body simulations. In this case, high-fidelity \textsc{Uchuu} lightcones are first constructed using sub-halo abundance matching (SHAM), with a single free scatter parameter calibrated to reproduce the observed two-point clustering. The resulting catalogues are then used to derive the halo occupation distribution (HOD) adopted to populate the lower-resolution GLAM $N$-body simulations. In terms of lightcone construction, the \textsc{Patchy} mocks are generated from a single cubic box of side length $2.5\,h^{-1}\,\mathrm{Gpc}$, whereas the \textsc{Uchuu} lightcones are assembled from simulation snapshots at different redshifts, thereby incorporating galaxy evolution along the lightcone, using boxes of side length $1\,h^{-1}\,\mathrm{Gpc}$.
The cosmology of the \textsc{Patchy} mocks is $\Omega_{\rm m}=0.307115$, $\Omega_{\rm b}=0.048206$, $\sigma_8=0.8288$, $n_s=0.9611$, and $h=0.6777$, while the \textsc{Uchuu} mocks adopt $\Omega_{\rm m}=0.309$, $\Omega_{\rm b}=0.0486$, $\sigma_8=0.816$, $n_s=0.9667$, and $h=0.677$. We use $2048$ realisations from \textsc{Patchy} and $2000$ realisations from \textsc{Uchuu} to estimate the covariance matrices. 

Although the \textsc{Uchuu} mocks were shown to provide a more accurate description of the BOSS two-point statistics~\cite{Ereza:2023zmz}, their performance for higher-order statistics, and in particular for the parity-odd kurto spectra studied here, has not yet been established. Moreover, ref.~\cite{Philcox:2024mmz} found that, for most BOSS subsamples, the \textsc{Uchuu} mocks exhibit a larger variance for the parity-odd 4PCF than the \textsc{Patchy} mocks, suggesting that previous parity analyses based on \textsc{Patchy} may have underestimated the sample variance. The low dimensionality of the kurto spectra allows us to quantify such differences directly by estimating the full covariance from both mock suites. While it is not possible to determine a priori which mock catalogue provides the more faithful description of the observed kurto spectra, comparing the results obtained with the two suites allows us to assess the robustness of our conclusions to the choice of mocks.  
The more detailed discussion of covariance matrices can be found in section~\ref{sec:BOSS-results}.

For both sets of mocks, the weighting scheme differs from that used for

the observed galaxy catalogues. For the \textsc{Patchy} mocks, both mock galaxies and randoms are weighted as
\begin{equation}
    w_{\mathrm{g}}^{\rm P} = w_{\mathrm{r}}^{\rm P} = w_{\rm cp}^{\rm P} \, w_{\rm veto} \,,
    \label{eq:weight-patchy}
\end{equation}
where fibre collisions are implemented in an approximate manner and accounted for by $w_{\rm cp}^{\rm P}$, while $w_{\rm veto}$ removes galaxies falling within the veto masks (e.g. bright stars and regions with poor imaging).\footnote{See \url{https://www.skiesanduniverses.org/page/page-3/page-15/page-9/} for a more detailed description of mock weighting.}
For the \textsc{Uchuu} mocks, the weighting scheme differs between the galaxy and random catalogues,
\begin{align}
    w_{\mathrm{g}}^{\mathrm{U}} &= w_{\rm cp} w_{\rm npcf} \, , \\
    w_{\mathrm{r}}^{\mathrm{U}} &= w_{\rm npcf} \, ,
\end{align}
where $w_{\rm npcf}$ is an extra weight to adjust the contribution of each sample in the combined catalogues \citep[LOWZ plus CMASS;][]{Ereza:2023zmz}.

%%%%%%%%%%%%%%%%%

In the case of BOSS, We approximate the normalisation factor defined in eq. \eqref{eq:I44} as a Monte Carlo sampling over the catalogue objects,
\begin{equation}
    I_{44} \approx \sum_{i=1}^{N_{\mathrm{g}}} w^3_{\mathrm{g}, \, i} \, n^3_{\mathrm{g}, \, i} \, w^4_{\mathrm{FKP}, \, i} \approx \alpha_{\mathrm{r}} \sum_{i=1}^{N_{\mathrm{r}}} w^3_{\mathrm{r}, \, i} \, n^3_{\mathrm{r}, \, i} \, w^4_{\mathrm{FKP}, \, i} \, ,
    \label{eq:I44_BOSS}
\end{equation}
where we follow previous work on BOSS and assume $w_{\mathrm{g}} \, n_{\mathrm{g}}$ the Poisson-distributed quantity.\footnote{If we were to consider the density alone as the Poisson-distributed quantity, the power on $w_{\mathrm{g}}$ would be $4$. We verified that this does not influence the final results.} We stress that for each sample (BOSS data, or different mocks), we use the corresponding definition of the galaxy or random systematic weights. In practice,  we often prefer to use the summation over the randoms as an approximation, as it is less noisy due to the larger density of the catalogue. For the power spectrum, where we use the $I_{22}$ as normalisation,\footnote{In general, we define 
$$I_{\rm NN} = \int \diff \mathbf{x} \, w_{\mathrm{FKP}}^N(\mathrm{x}) \, [w_{\mathrm{g}}(\mathrm{x}) \, n_{\mathrm{g}}(\mathrm{x})]^N \, .$$} the difference between the use of the galaxies or the randoms in the approximation is within $5\%$. However, we found that in the trispectrum case, these two methods can differ by more than $50\%$. In this paper, we use the estimation from the galaxy catalogue.

\subsection{DESI-DR1 data and mocks}\label{subsec:desi_data}

The second dataset that we utilise is the full LRG sample of DESI-DR1 \citep{desi-dr1,Ross:2024nlf}. This is the same sample analysed in the PV 4PCF by refs.~\cite{Slepian:2025kbb,Hou:2025cey}. The DESI-DR1 LRG sample covers the redshift range $0.4 < z < 1.1$ with a number density of $n(z) \sim 5\times 10^{-4} \, h^3 \, \mathrm{Mpc}^{-3}$ up to $z \sim 0.8$, see figure~\ref{fig:nz_comparison} for a comparison beween BOSS DR12 and DESI DR1 $n(z)$. 
Similar to BOSS-DR12, DESI-DR1 data are divided into two fields of view in the North and South Galactic Caps (also called NGC and SGC), which cover a sky area of $3650 \, \mathrm{deg}^2$ and $2089 \, \mathrm{deg}^2$, respectively \cite{DESI:2024aax}.  The fiducial cosmology we assume for our measurements is the same as that of the DESI Collaboration 
with: $h=0.6736$, $\Omega_{\rm CDM} = 0.26447$, $\Omega_{\rm b} = 0.0493$, $A_s=2.083\times 10^{-9}$, and $n_s=0.9649$.

For DESI-DR1, the weighting scheme is the same for data and mocks, and for galaxy and random catalogues. The collaboration directly provides a total weight for each object,
\begin{equation}
    w_{\mathrm{g}}^{\mathrm{DESI}} = w_{\mathrm{r}}^{\mathrm{DESI}} = w_{\mathrm{tot}} \, ,
\end{equation}
which captures all the systematic effects after correcting for the up-weighting of the completeness due to overlapping tiles~\cite{Ross:2024nlf}. Additionally, due to the low completeness of the DR1 sample, the number density is provided as a function of both redshift and tile number, $n(x) = n(z, N_{\mathrm{tile}})$.\footnote{The total weight and number density columns are called \texttt{WEIGHT} and \texttt{NX} in the catalogues.} 
This is the quantity we use for the computation of the FKP weights, for which we also use the same $P_0$ used in the BOSS-DR12 analysis \cite{DESI:2024aax}.

For the covariance estimation, the DESI Collaboration provides three sets of mock catalogues. The first set is produced with the Effective Zel’dovich approximation method~\cite{Chuang:2014vfa,eBOSS:2020wwo} and refereed to as \textsc{EZmock}. This algorithm allows the production of a large number of mocks with large volumes. The collaboration provides $1000$ \textsc{EZmock} cut from boxes of $(6 \, h^{-1} \, \mathrm{Mpc})^3$, which contain the whole survey volume without the need for tiling. To produce these mocks, as running the fibre assignment pipeline for a $1000$ mocks would have a high computational cost, an approximate fibre assignment algorithm is used, namely the fast fibre assignment \citep[FFA;][]{Bianchi:2025JCAP}; therefore, there will be small-scale differences between these mocks and the data due to the approximate simulation algorithm and the incorrect fibre assignment. These mocks are used for numerical covariance estimations. The other two sets of mocks are produced from \textsc{AbacusSummit} simulations~\cite{Garrison:2018juw,Maksimova:2021ynf}. These are $N$-body simulations. The drawback of these mocks is their limited number and their size; there are only $25$ realisations; their volume is $(2 \, h^{-1} \, \mathrm{Mpc})^3$, and the box has to be replicated to cover the survey volume. The difference between the two Abacus mocks is the fibre assignment algorithm. The first set uses the FFA algorithm, so it enables understanding the effect of box replication and possibly the effect of approximate simulations, especially for higher order statistics. The second set of mocks uses the official fibre assignment algorithm, the alternative merged target ledgers \citep[altMTL;][]{desi-altMTL}. Making this mocks the one with systematic effects closest to real data.

We follow the choice of DESI collaboration for the normalisation of the power spectrum, $I_{22}$, and extend it to that of trispectrum, and hence kurto spectra, $I_{44}$. This choice differs from that of BOSS in that the integral is not Monte Carlo sampled, but it is approximated as a summation over the grid~\cite{DESI:2024aax}. Hence, the power spectrum normalisation reads as
\begin{equation}
    I_{22} \approx \frac{\alpha_{\mathrm{r}}}{\diff V_{\rm cell}} \sum_i N_{\mathrm{d}}(\mathbf{x}_i) N_{\mathrm{r}}(\mathbf{x}_i) \, ,
    \label{eq:I22_desi}
\end{equation}
where $\diff V_{\rm cell}$ is the cell volume for each mesh grid, computed as the total box volume divided by the total number of mesh grids, and $N_{\mathrm{d}}(\mathbf{x}_i)$ and $N_{\mathrm{r}}(\mathbf{x}_i)$ are the weighted density over the grid. Using the same approximation for the trispectrum case, we have
\begin{equation}
    I_{44} \approx \frac{\alpha^\lambda_{\mathrm{r}}}{\diff V^3_{\rm cell}} \sum_i N^{4-\lambda}_{\mathrm{d}}(\mathbf{x}_i) N^{\lambda}_{\mathrm{r}}(\mathbf{x}_i) \, , 
    \label{eq:I44_desi}
\end{equation}
where $\lambda=\{0,1,2,3,4\}$. We use $\lambda=3$ as it gives the most stable results in our tests. 

Finally, following ref.~\cite{Hou:2025cey}, we also perform a cross-correlation analysis of sub-patches of the two fields of view, NGC and SGC (see section~\ref{sec:chicross}). We report the cuts strategy in table~\ref{tab:selection}. 
\begin{table}
  \centering
  \begin{tabular}{cccccc}
    \toprule
    Cut & NGC-Full & SGC-Full & NGC-1 & NGC-2 & SGC-3 \\
    \midrule
    RA & All & All & $(110, 260)$ & $(180, 260)$ & All \\
    Dec & All & All & $(-10, 8)$ & $(30, 40)$ & All \\
    $z$ & $(0.4, 1.1)$ & $(0.4, 1.1)$ & $(0.4, 0.8)$ & $(0.4, 0.8)$ & $(0.4, 0.8)$ \\
    \bottomrule
  \end{tabular}
  \caption{The Definition of the DESI LRG sub-patches.}
  \label{tab:selection}
\end{table}

\subsection{The null hypothesis test} \label{sec:chi2}

By construction, the two kurto spectra defined in eqs.~\eqref{eq:P_2x2} and \eqref{eq:P_3x1} have non-zero expectation values only in the presence of a parity-violating signal. Since the mock catalogues do not contain any cosmological PV signal, a non-zero measurement in the mocks would indicate parity-odd systematics. Such systematics are not generally expected to be significant; see ref.~\cite{Hou:2023MNRAS} for a detailed discussion.

We start from the null-hypothesis that the data does not have a cosmological PV signal. Then, a significant deviation from zero of the data vector would reject the null hypothesis and suggest the existence of a PV signal in the data. Additionally, we note that this kind of hypothesis test is independent of the theoretical template of PV. 

In order to verify the null-hypothesis, since the kurto spectra data vector is noisy (where the noise depends on both cosmic variance and stochasticity~\cite{Gao:2025yqd}), we compare the chi-square statistics of data and mocks. If the data were to have a significantly larger chi-square than the mocks, we would reject the null hypothesis, and vice versa. We define the chi-square as
\begin{equation}
\label{eq:chi_square}
    \chi^2 = \xi^{\rm T} \mathbf{C}^{-1} \xi \, ,
\end{equation}
where $\xi$ is the data vector ($\mathcal{P}_{2\times 2}$ or $\mathcal{P}_{3\times 1}$) measured from data and mocks, while $\mathbf{C}$ is the true covariance matrix. As mentioned earlier, we use the numerical covariance matrices estimated from the available suites of mocks corresponding to each datasets. Therefore, the numerical covariance matrices are computed as
\begin{align}
    \mathrm{C}_{ij}^{\mathrm{N}} &= \langle \left[\xi(k_i) - \langle \xi(k_i) \rangle \right]^{\rm T} \, \left[\xi(k_j) - \langle \xi(k_j) \rangle \right] \rangle \nonumber\\
    &= \frac{1}{N_{\mathrm{s}}-1}\sum_{a=1}^{N_{\mathrm{s}}} \left[\xi(k_i) - \bar{\xi}(k_i)\right]_a \, \left[\xi(k_j) - \bar{\xi}(k_j)\right]_a \, ,
    \label{eq:covariance}
\end{align} 
where we have the ensemble average defined for eq.~\eqref{eq:P_2x2}, the subscript $a$ is the mock index, $N_{\mathrm{s}}$ is the number of mocks, and $\bar{\xi}(k_i)$ is the average of the data vector defined as
\begin{equation}
    \bar{\xi}(k_i) = \frac{1}{N_{\mathrm{s}}}\sum_{a=1}^{N_{\mathrm{s}}} \xi_a(k_i) \, .
    \label{eq:d_average}
\end{equation}
To take into account the non-Gaussianity of the covariance induced by the limited number of mocks used in its estimation, we multiply the covariance matrix by the Wishart factor \citep[also known as Hartlap factor;][]{2007A&A...464..399H}, 
\begin{equation}
    \hat{\mathrm{C}}_{ij} = \frac{N_{\mathrm{s}} - 1}{N_{\mathrm{s}} - N_{\mathrm{b}} - 2} \, \mathrm{C}_{ij}^{\mathrm{N}} \, ,
    \label{eq:covariance_general}
\end{equation}
where $N_{\mathrm{b}}$ is the dimension of the data vector. 

We stress once again that the use of numerical covariance constitutes one of the main advantages of the kurto spectra analysis. As mentioned earlier, the previous PV analyses based on the 4PCF~\cite{Hou:2024blc,Philcox:2024mmz,Slepian:2025kbb,Hou:2025cey}, relied on analytic covariance assuming a Gaussian random field~\cite{Hou:2021ncj}. For DESI and BOSS, the number of available mocks is $N_{\mathrm{s}} = 1000$ and $2048$, respectively, while the 4PCF is a much larger data vector $N_{\mathrm{b}} \sim 2760$ and $18\,768$ for $10$ and $18$ radial bins. Thus, the numerical covariance is not robust for this estimator and one is forced to use analytic covariances. However, the analytic covariance based on Gaussian assumption suffers from the inaccuracy and the mis-evaluation of several components, including anisotropy due to redshift-space distorsion (RSD) in correlations functions, contribution from connected higher order statistics, convolution with survey geometry. These usually result in an underestimation of the noise, as emphasized in ref.~\cite{Krolewski:2024JCAP}. 

In contrast, for each of the two kurto spectra, the $N_{\mathrm{b}}$ is orders of magnitude smaller than the number of mocks, as they have the dimensionality of a power spectrum. In our study cases, we have $N_{\mathrm{b}} = 70$ and $N_{\mathrm{b}} = 32$ for BOSS-DR12 and DESI-DR1, respectively. Therefore, assuming that the mocks are a fair representation of data in terms of the applied HOD model and survey realism, the numerical covariance estimate is robust and accurate, and captures all the non-Gaussian information.

\subsection{The signal sensitive cross statistics}\label{sec:chicross}

If the mock's covariance is not accurate in probing the underlying data covariance, the resulting data $\chi^2$ will be discrepant with the mock $\chi^2$ distribution, even in the absence of cosmological PV signal. This effect has been referred to as `data-mock mismatch' in previous works. In ref.~\cite{Krolewski:2024JCAP}, a new statistic was constructed to circumvent the data-mock mismatch, namely the \textit{cross-chi-square}, defined as
\begin{equation}
    \chi^2_{\times} \equiv  \frac{1}{N_{\mathrm{p}}(N_{\mathrm{p}}-1)} \sum_{\mu \neq \nu} \xi^\mu_{a} (\hat{\mathbf{C}}^{-1})^{ab} \xi^\nu_{b} \, 
    \label{eq:chi_cross}
\end{equation}
where $N_{\mathrm{p}}$ is the number of patches, while $\mu$ and $\nu$ are the patch indices. 

In the absence of such a shared signal, and neglecting residual systematics or additional data--mock mismatch, $\chi^2_{\times}$ is expected to be centered around zero. The insensitivity of this statistic to data-mock mismatch is shown in~\ref{append:chi-cross}. 

We note, however, that in practice a discrepancy in $\chi^2_{\times}$ between data and mocks does not necessarily imply a cosmological PV signal, since it may also arise for at least three reasons:
\begin{itemize}
    \item a genuine cosmological PV signal shared across the patches; 
    \item a mismatch in the variance of the observable between data and mocks, such that even if the parity-odd observable itself has zero mean under the null hypothesis, the mocks may underestimate its scatter. The resulting mock $\chi^2_{\times}$ distribution is then too narrow, and the observed $\chi^2_{\times}$ can appear artificially discrepant. This is another form of data--mock mismatch~\cite{Krolewski:2024JCAP};
    \item correlated features between patches induced by systematics that are not captured by the mocks, in which case the data $\chi^2_{\times}$ distribution need not be centred at zero ~\citep[see also discussion in section 4.5 of ref.][]{Krolewski:2024JCAP}.
\end{itemize}

It is important to emphasise that the cross-chi-square defined in eq.~\eqref{eq:chi_cross} is actually a data compression rather than a true chi-square statistic as what we defined in eq.~\eqref{eq:chi_square}. For this reason, the covariance that appears in $\chi_\times^2$ can be arbitrary. As long as we use the same covariance for both data and mocks, their difference (consistency) will inform the existence (non-existence) of a cosmological PV signal. We will specify the covariance when describing the results of the analysis for each survey. Note that, in the main text, we take $N_{\mathrm{p}}=2$, where we demonstrate the cross-correlation of each combination of two patches. 

Ref.~\cite{Krolewski:2024JCAP} defined another statistic, $\chi^2_{\rm null}$, which isolates the data-mock mismatch in the $\chi^2$ analyses. This contribution is expected to be nonzero if the parity-even eight-point correlation function of the data differs from the mocks. We do not perform the analyses in the paper, since there is no statistically significant detection of PV found in the null hypothesis test or in the signal-sensitive cross-chi-sqaure statistics. 

\section{Results}\label{sec:results}

In this section, we present the results from BOSS-DR12 and DESI-DR1 analyses. For each dataset, we discuss the measured kurto spectra, $\mathcal{P}_{2\times 2}$ and $\mathcal{P}_{3\times 1}$, the correlation matrix, the chi-square analyses, and the cross-chi-square results. For BOSS, we show the cross-chi-square analysis only for the two full patches, NGC and SGC, whereas for DESI we perform a more detailed study that also includes sub-patches within the NGC.

\begin{figure}[t]
    \begin{centering}
    \begin{subfigure}{\textwidth}
    \includegraphics[width=0.9\linewidth]{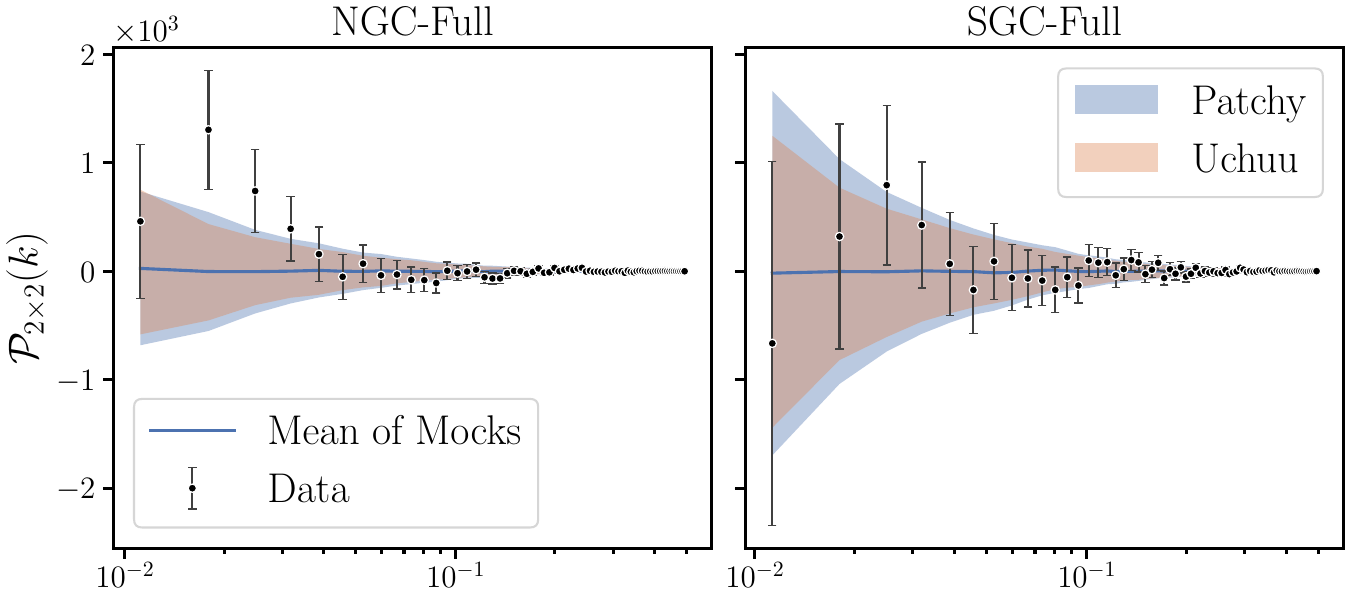}
    \end{subfigure}\vspace{0.1in}
    \begin{subfigure}{\textwidth}
    \includegraphics[width=0.9\linewidth]{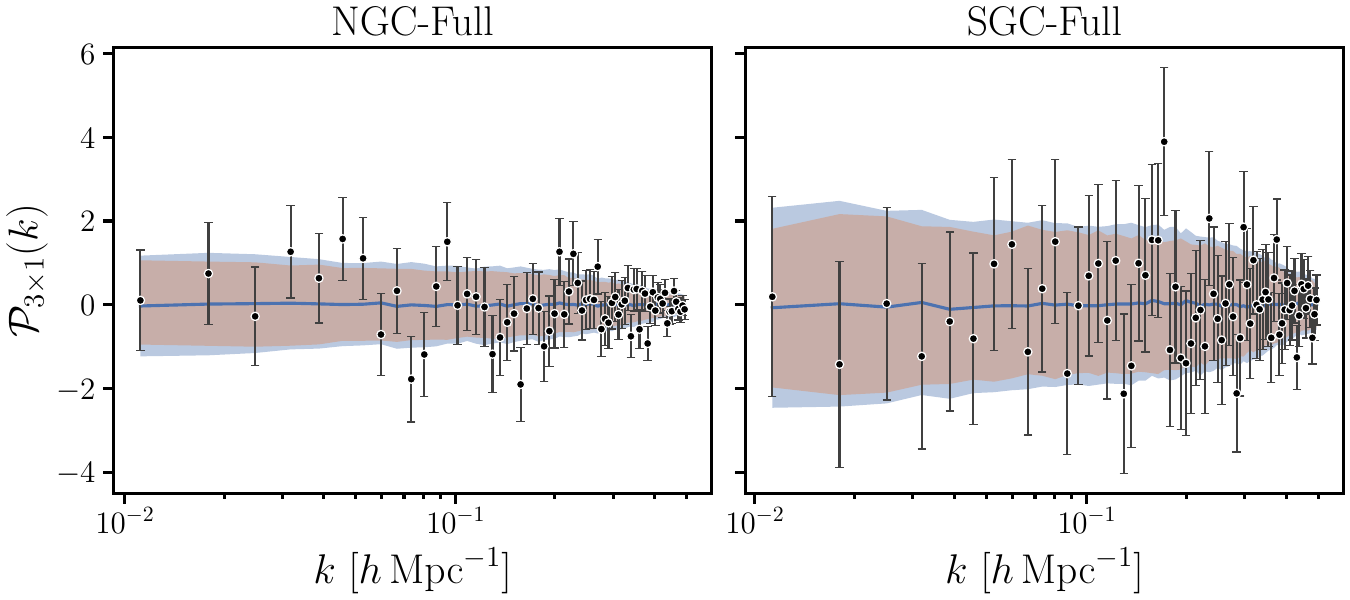}
    \end{subfigure}
    \end{centering}
    \caption{Kurto spectra for BOSS-DR12 LRG sample. The black dots are the data measurements. The vertical error bars are estimated from the \textsc{Patchy} mocks. The blue solid line is the \textsc{Patchy} mock average, and the blue and orange shaded areas show $1\sigma$ spread of the \textsc{Patchy} and \textsc{Uchuu} mocks. The top and bottom rows show the $\mathcal{P}_{2\times2}$ and $\mathcal{P}_{3\times1}$ kurto spectra, respectively, of the NGC (left column) and SGC (right column) full fields of view 
    }
    \label{fig:P2x2_P3x1_BOSS_data}
\end{figure}

\subsection{BOSS-DR12} \label{sec:BOSS-results}

\begin{figure}
    \centering
    \begin{subfigure}{0.93\textwidth}
    \includegraphics[width=\linewidth]{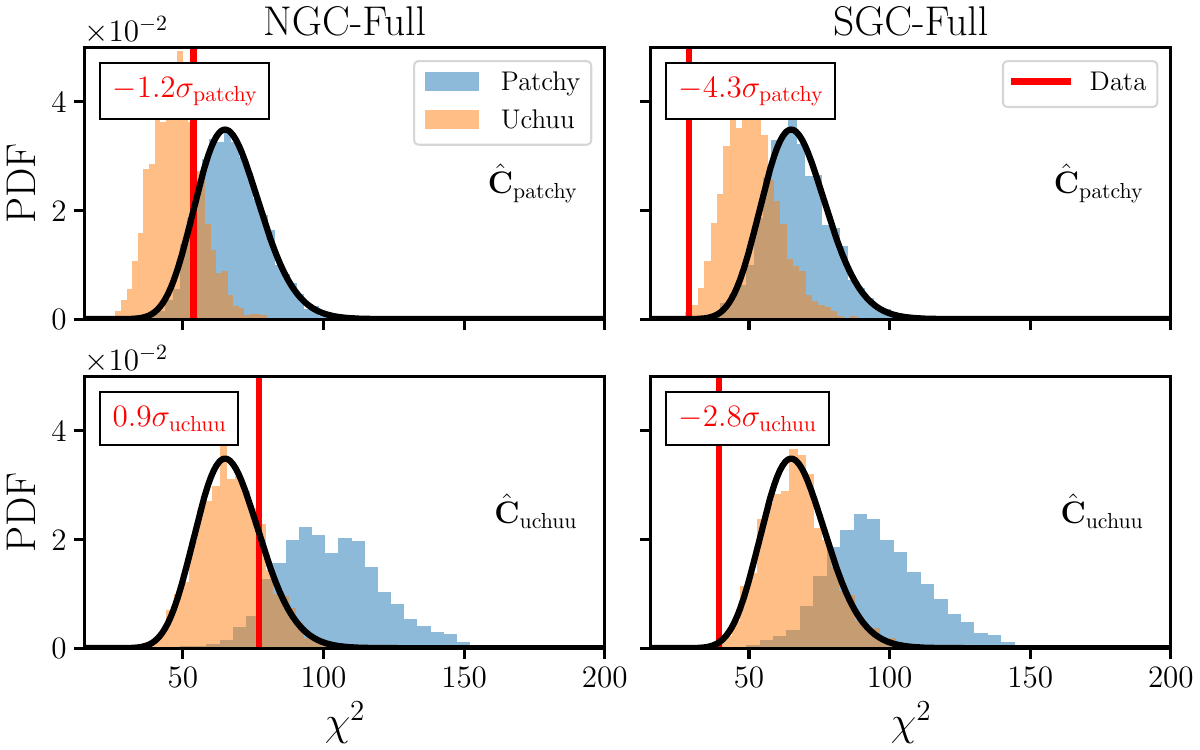}
    \caption{Chi-square for the $\mathcal{P}_{2\times 2}$ kurto spectra.}
    \label{subfig:chi2_P2x2_BOSS}
    \end{subfigure}\vspace{0.15in}
    \begin{subfigure}{0.93\textwidth}
    \includegraphics[width=\linewidth]{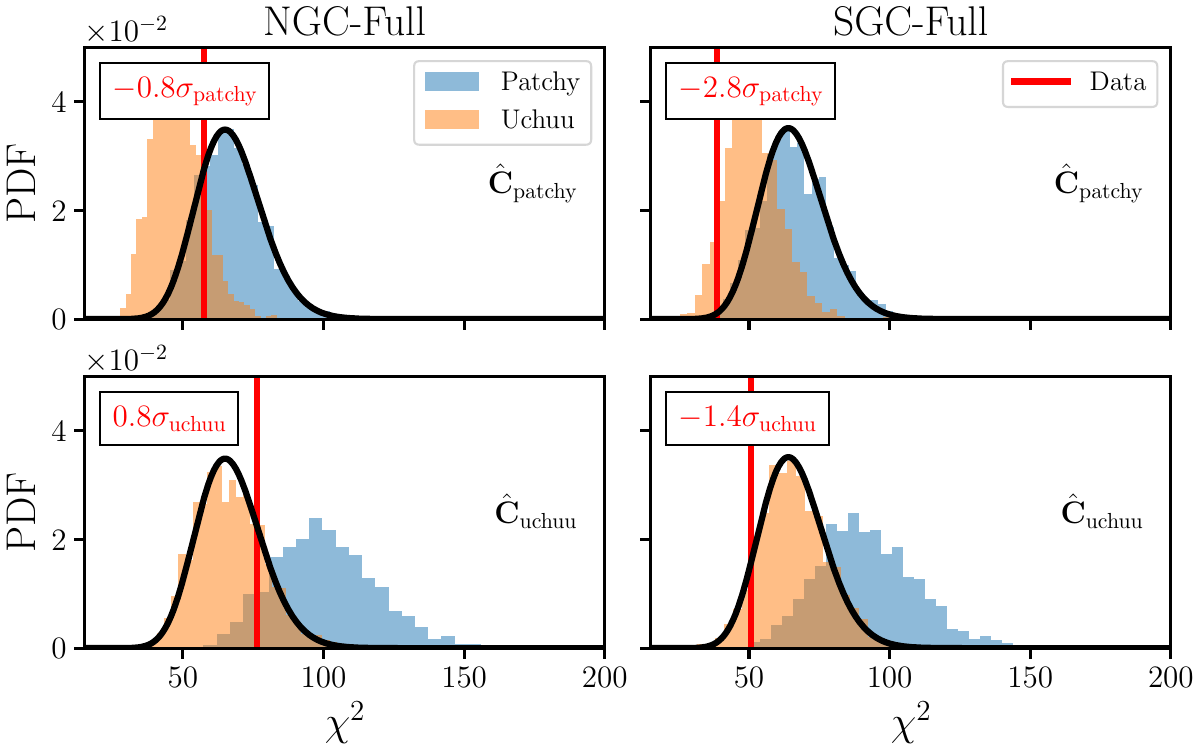}
    \caption{Chi-square for the $\mathcal{P}_{3\times 1}$ kurto spectra.}
    \label{subfig:chi2_P3x1_BOSS}
    \end{subfigure}
    \caption{The $\chi^2$ distribution from BOSS-DR12. For both sub-figures, in the first row use the covariance from \textsc{Patchy}, while in the second row the one from \textsc{Uchuu} mocks. The data $\chi^2$ is presented with a solid red vertical line, while the distributions of \textsc{Patchy} and \textsc{Uchuu} data vectors are shown by the blue and orange bars. The solid black curve represents the theoretical chi-square distribution for the given degrees of freedom (d.o.f), where we found the numerically fitted one matches the theoretical prediction using Wishart correction.}
    \label{fig:chi2_P2x2_P3x1_BOSS}
\end{figure}
In figure~\ref{fig:P2x2_P3x1_BOSS_data}, we show the measured $\mathcal{P}_{2\times2}$ (top) and $\mathcal{P}_{3\times1}$ (bottom) kurto spectra for the BOSS-DR12 LRG sample in the NGC (left) and SGC (right). The black points show the measurements from the data, with error bars derived from the diagonal elements of the covariance matrix estimated from the \textsc{Patchy} mocks. The blue lines show the mean of the \textsc{Patchy} mocks, while the blue and orange shaded regions indicate the $1\sigma$ scatter of the \textsc{Patchy} and \textsc{Uchuu} mocks, respectively.

Overall, both parity-odd kurto spectra fluctuate around zero, with most of the measurements lying within the scatter expected from the mocks. Although a few large-scale points in $\mathcal{P}_{2\times2}$ lie outside the plotted $1\sigma$ range, this should not be interpreted as evidence for a PV signal. If the apparent signal were sourced by a cosmological PV signature, one would expect a consistent contribution across different patches of the survey, up to statistical fluctuations. By contrast, a feature confined to a single patch may instead reflect a local fluctuation or an unmodelled systematic effect. This motivates the null-hypothesis test based on $\chi^2$ statistic and the signal-sensitive cross-patch test based on $\chi^2_{\times}$.  

It is worth noting that the mock mean is more consistent with zero than the data, indicating that no apparent PV signal is generated within the mock catalogues. At the same time, the overall agreement between the scatter of the mocks and that of the data suggests that the measured fluctuations are still dominated by parity-even noise, as discussed in ref.~\cite{Gao:2025yqd}. In other words, any possible parity-odd signal remains subdominant to the parity-even noise of the survey. We find a similar behaviour for DESI-DR1 (see figures~\ref{fig:P2x2_DESI_data_all_patches} and~\ref{fig:P3x1_DESI_data} in section~\ref{subsec:DESI_DR1}).

In figure~\ref{fig:chi2_P2x2_P3x1_BOSS}, we show the $\chi^2$ statistics for the BOSS-DR12 data (red line) and the \textsc{Patchy} and \textsc{Uchuu} mocks (blue and orange distributions). Sub-figures~\ref{subfig:chi2_P2x2_BOSS} and \ref{subfig:chi2_P3x1_BOSS} correspond to the $\mathcal{P}_{2\times2}$ and $\mathcal{P}_{3\times1}$ kurto spectra, respectively. In each subfigure, the first row shows the $\chi^2$ values computed using the covariance estimated from the \textsc{Patchy} mocks, while the second row uses the covariance estimated from the \textsc{Uchuu} mocks.

For both kurto spectra, the data $\chi^2$ values do not lie in the high-$\chi^2$ tail of the mock distributions, and we therefore find no evidence for a non-zero PV signal in BOSS-DR12. In most cases, the data $\chi^2$ lies on the left-hand side of the \textsc{Patchy} mock distribution, indicating that the observed kurto spectra are closer to zero than a typical \textsc{Patchy} realisation. Since the mock catalogues have a mean parity-odd kurto spectrum consistent with zero (see figure~\ref{fig:P2x2_P3x1_BOSS_data}), their non-zero realisations quantify the null fluctuations of the estimator. The offset between the data and \textsc{Patchy} mock $\chi^2$ distributions therefore indicates that the parity-odd kurto spectra exhibit smaller null fluctuations in the BOSS data than in the \textsc{Patchy} mocks. At the same time, the data $\chi^2$ values are generally more consistent with the \textsc{Uchuu} distributions than with those of \textsc{Patchy}, which may reflect the higher-fidelity construction of the \textsc{Uchuu} mocks~\cite[full versus approximate $N$-body and different galaxy assignment model with fewer free parameters;][]{Ereza:2023zmz}. The only case in which the data $\chi^2$ lies on the right-hand side of the corresponding mock distribution is the NGC-Full result with the \textsc{Uchuu} covariance, but the deviation is below $1\sigma$.

The mock $\chi^2$ distributions computed with their corresponding covariance matrices are in good agreement with the theoretical $\chi^2$ expectation. For BOSS-DR12, we use $N_{\mathrm b}=70$ $k$-bins. Since both the \textsc{Patchy} and \textsc{Uchuu} suites contain at least $2000$ realisations, the Wishart correction factor is $\sim 0.965$, giving an effective number of degrees of freedom of about 67. Naturally, if the covariance from the other mock suite is used, the mock $\chi^2$ distributions deviate from the corresponding theoretical $\chi^2$ curve. The \textsc{Uchuu} mock realisations generally yield smaller $\chi^2$ values than the \textsc{Patchy} realisations, suggesting that the \textsc{Uchuu} kurto spectra are more tightly distributed around zero for this statistic.

\begin{figure}
    \centering
    \begin{subfigure}{0.92\textwidth}
    \includegraphics[width=\linewidth]{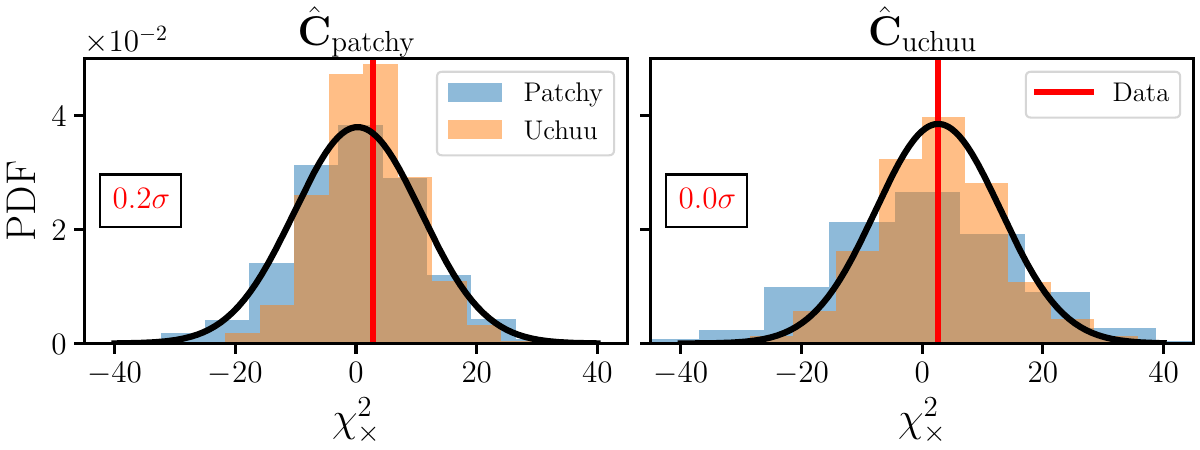}  
    \caption{Cross-chi-square for the $\mathcal{P}_{2\times 2}$ kurto spectra.}
    \label{subfig:chi_cross_P2x2_BOSS}
    \end{subfigure}\vspace{0.15in}
    \begin{subfigure}{0.92\textwidth}
    \includegraphics[width=\linewidth]{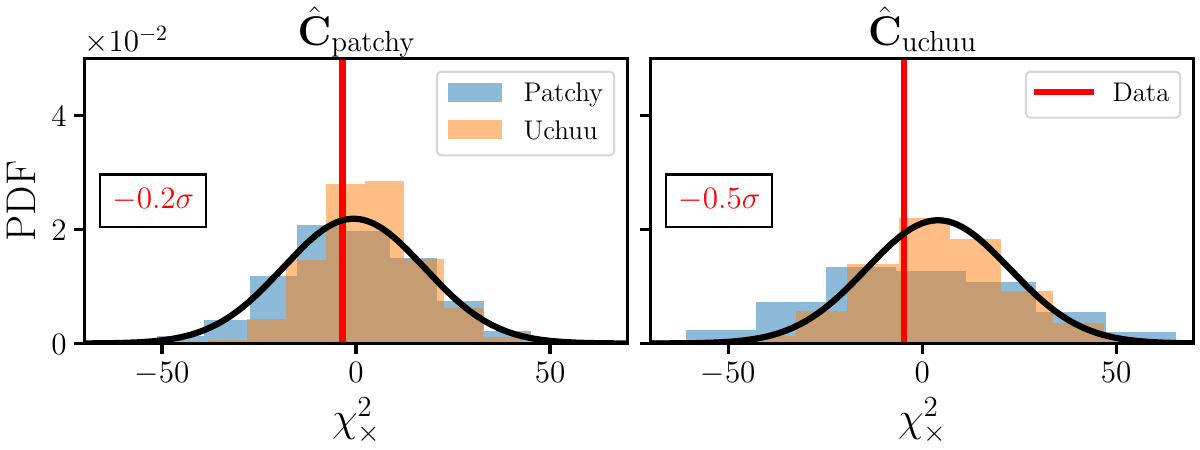}  
    \caption{Cross-chi-squared for the $\mathcal{P}_{3\times 1}$ kurto spectra.}
    \label{subfig:chi_cross_P3x1_BOSS}
    \end{subfigure}
    \caption{The $\chi_\times^2$ distribution between NGC and SGC for $\mathcal{P}_{2\times 2}$ (top) and $\mathcal{P}_{3\times 1}$ (bottom) measurements from BOSS-DR12. For the left panel, the covariances are estimated from \textsc{Patchy}, while for the right panel, we use the covariance from \textsc{Uchuu} mocks. We show the data $\chi^2_\times$ with a red vertical line, while the distributions of \textsc{Patchy} and \textsc{Uchuu} data vectors are shown by the blue and orange histograms. The black curve represents a Gaussian distribution centred on $0$ and standard deviation computed from the mocks whose covariance we use.}
    \label{fig:chi_cross_P2x2_P3x1_BOSS}
\end{figure}

We present the $\chi_\times^2$ data compression in figure~\ref{fig:chi_cross_P2x2_P3x1_BOSS}. Sub-figures~\ref{subfig:chi_cross_P2x2_BOSS} and \ref{subfig:chi_cross_P3x1_BOSS} show the cross-chi-square for the $\mathcal{P}_{2\times 2}$ and $\mathcal{P}_{3\times 1}$ kurto spectra, respectively. In each sub-figure, the left and right panels show the results obtained from the NGC-Full covariance of the \textsc{Patchy} and the \textsc{Uchuu} mocks, respectively. We use this statistic to further confirm the null-hypothesis. The $\chi_\times^2$ follows a Gaussian distribution nicely, while the data vector is always consistent with zero, showing that there is no shared feature in NGC-Full and SGC-Full data.

\begin{figure}
    \centering
    \includegraphics[width=1.\linewidth]{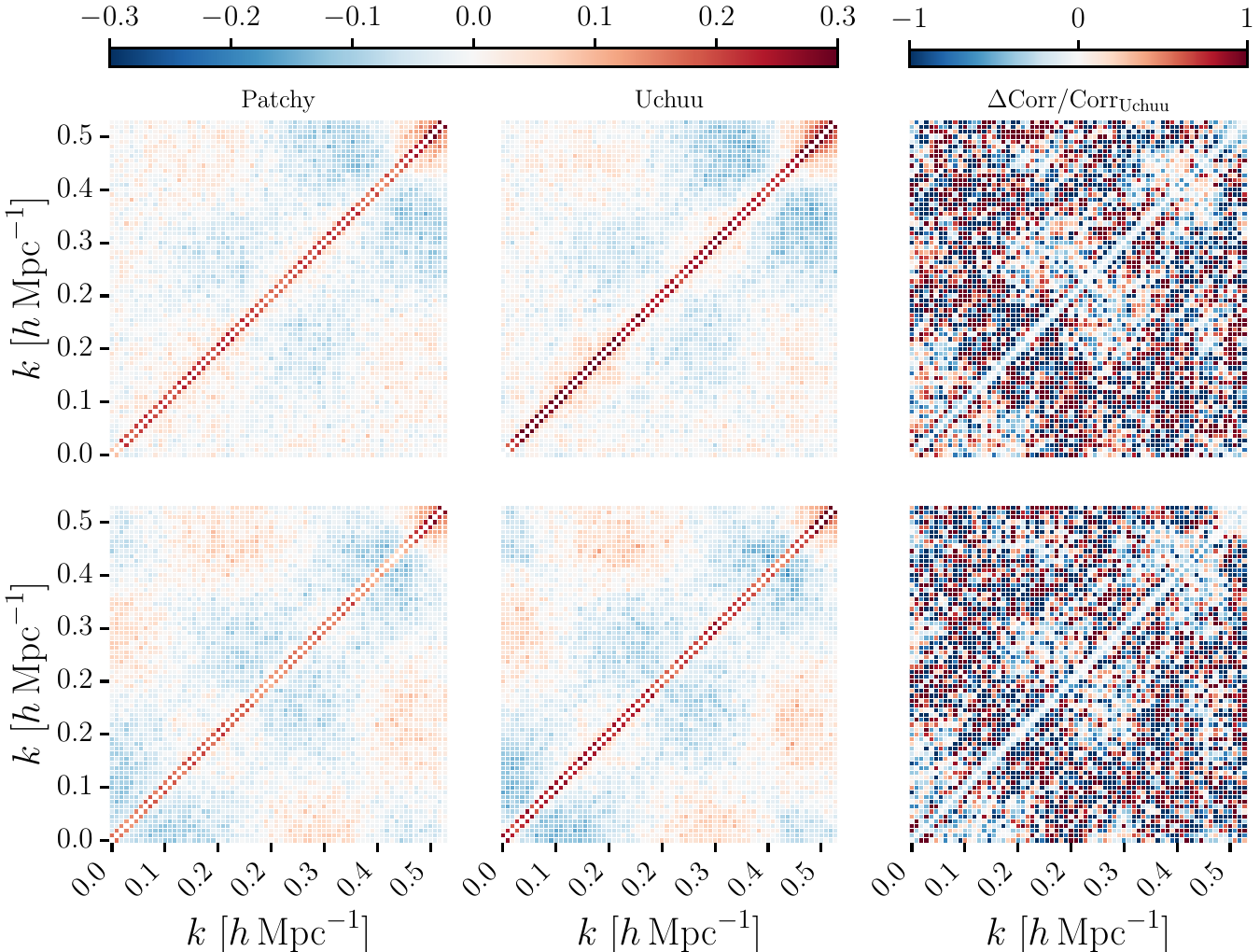}
    \caption{Correlation matrices for BOSS DR12 NGC. The top row shows the matrices for $\mathcal{P}_{2\times 2}(k)$, and the bottom one for $\mathcal{P}_{3\times 1}(k)$. In the left, middle, and right columns we present the \textsc{Patchy}, the \textsc{Uchuu} correlation matrix, and relative difference, respectively. We note that the colorbar range of the first two columns is set to be $[-0.3, 0.3]$ to demonstrate the off-diagonal structure, while the diagonal terms are strictly one. The colorbar of the last column is $[-1,1]$. In all panels, we mask out the diagonal terms.}
    \label{fig:correlation-matrix}
\end{figure}

To better understand the $\chi^2$ and $\chi_\times^2$ results, we show in figure~\ref{fig:correlation-matrix} the correlation matrices of $\mathcal{P}_{2\times 2}$ (top row) and $\mathcal{P}_{3\times 1}$ (bottom row). The left and middle columns show the matrices estimated from \textsc{Patchy}, \textsc{Uchuu} mocks, while the right column shows their relative difference. Since the SGC exhibits similar correlation structure, we show only the results from NGC. The correlation matrices of both estimators are close to diagonal with only weak off-diagonal correlations. This is consistent with the findings in a noise-dominated regime, where different $k$-bins of the kurto spectra are only weakly correlated~\cite{Gao:2025yqd}. We have verified that retaining only the diagonal elements of covariance matrix does not change the $\chi^2$ results. By contrast, in the PV signal-dominant regime, the parity-odd kurto spectra covariance is generally expected to have stronger off-diagonal structure, by analogy with related composite-field statistics such as skew spectra, for which the full covariance matrix is known to be important in forecasted cosmological constraints~\cite{MoradinezhadDizgah:2019xun,Hou:2022rcd}. The two estimators also exhibit somewhat different correlation patterns. This is consistent with the behaviour found in ref.~\cite{Gao:2025yqd}, where $\mathcal{P}_{2\times2}$ and $\mathcal{P}_{3\times1}$ were shown to have different spectral shapes, indicating sensitivity to different combinations of scales and hence different covariance structures.

\begin{figure}
    \centering
    \includegraphics[width=0.92\linewidth]{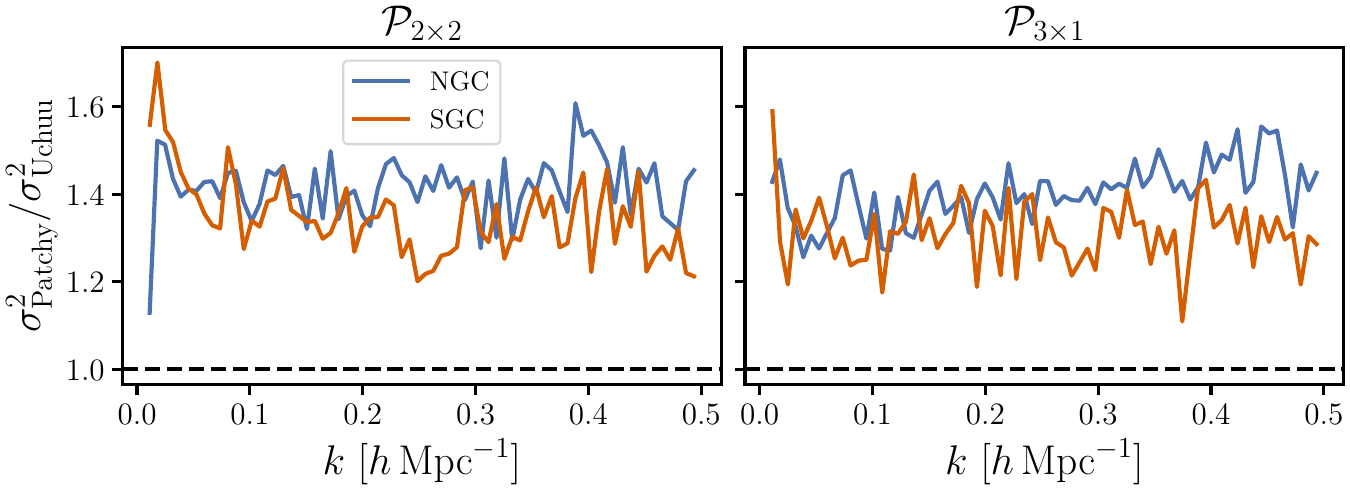}
    \caption{The ratio between of the variances estimated from \textsc{Patchy} and \textsc{Uchuu} mocks. The blue and red solid lines correspond to NGC and SGC, respectively. The left and right panels are the measurements from $\mathcal{P}_{2\times 2}$ and $\mathcal{P}_{3\times 1}$ kurto spectra.}
    \label{fig:variance-compare}
\end{figure}

These results further highlight one of the main advantages of the kurto-spectrum approach. Because the covariance matrix of the kurto spectra is close to diagonal, and because the dimensionality of the estimator is much lower than that of the full 4PCF, the covariance can be estimated numerically from mocks. This makes it possible to incorporate, in a natural way, effects that are difficult to capture in the analytic covariance treatments adopted in previous 4PCF analyses, such as departures from Gaussianity, anisotropies induced by redshift-space distortions, and survey-window effects; see ref.~\cite{Philcox:2022PRD} for a detailed discussion. These effects can contribute non-trivially even to the diagonal covariance terms. In particular, ref.~\cite{Philcox:2022PRD} reported an approximately factor-of-two difference between the diagonal elements of the numerical and analytic 4PCF covariances, illustrating the limitations of the analytic approximation. By contrast, the weak off-diagonal structure of the kurto-spectrum covariance and the minor impact of these terms on the inferred $\chi^2$ values indicate that this estimator is statistically much better behaved than the full 4PCF, which is another practical advantage of working with the compressed observable.

Although the \textsc{Patchy} and \textsc{Uchuu} correlation matrices are qualitatively similar, their relative difference exceeds $100\%$ for many off-diagonal elements. This should be interpreted with some care, since the off-diagonal correlations themselves are small, so even modest absolute differences can correspond to large relative changes. To compare the two set of mocks more directly, we show in figure~\ref{fig:variance-compare} the ratio of the diagonal covariance elements, i.e., the variance. We find that the \textsc{Patchy} variance is typically about $1.4$ times larger than that of \textsc{Uchuu}. This explains the behaviour seen in figure~\ref{fig:chi2_P2x2_P3x1_BOSS}: because the \textsc{Uchuu} kurto spectra exhibit smaller noise, the $\chi^2$ values of \textsc{Uchuu} realisations evaluated with the \textsc{Patchy} covariance (upper rows of figure~\ref{fig:chi2_P2x2_P3x1_BOSS}) lie on the left-hand side of the expected $\chi^2$ distribution. Interestingly, this trend differs from that found for the parity-odd 4PCF in ref.~\cite{Philcox:2024mmz}, where the \textsc{Uchuu} variance was reported to be larger than that of \textsc{Patchy}. This indicates that the relative behaviour of the two mock suites is statistic-dependent, and need not be the same for Fourier-space and configuration-space observables. In the original \textsc{Uchuu} study~\cite{Ereza:2023zmz}, the variance of the \textsc{Uchuu} two-point correlation function was also found to be larger than that of \textsc{Patchy}, while for the power spectrum the behaviour is scale-dependent: on large scales ($k<0.15\,h\,\mathrm{Mpc}^{-1}$), \textsc{Uchuu} has about $10\%$ larger variance than \textsc{Patchy}, whereas on smaller scales ($k>0.15\,h\,\mathrm{Mpc}^{-1}$) its variance becomes smaller, with the relative difference growing from about $10\%$ to $30\%$ toward higher $k$. For the kurto spectra, because mode coupling is involved, the variance ratio cannot be inferred straightforwardly from the corresponding power-spectrum ratio alone.

\subsection{DESI-DR1} \label{subsec:DESI_DR1}

\begin{figure}
    \centering
    \begin{subfigure}{0.92\textwidth}
    \includegraphics[width=\linewidth]{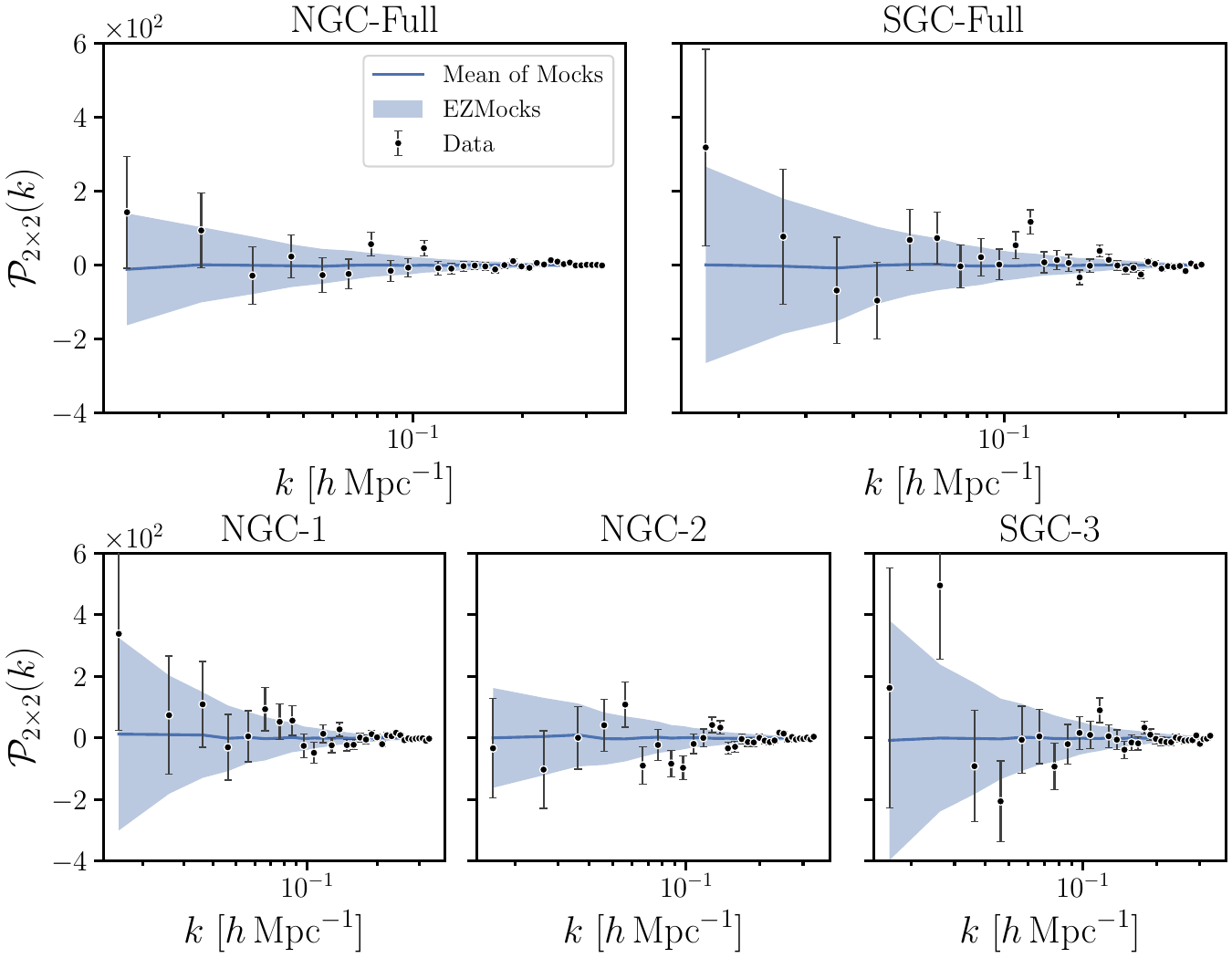}
    \end{subfigure}
    \caption{Measured $\mathcal{P}_{2\times 2}$ from DESI-DR1 data. The black dots are the data points, the error bars of which we estimated from the \textsc{EZmocks}. The blue solid line is the average of \textsc{EZmocks}, and the blue shaded area is their $1 \sigma$ error (which is the same error we show for the data).} 
    \label{fig:P2x2_DESI_data_all_patches}
    
    \vspace{0.3in}

    \centering
    \begin{subfigure}{0.92\textwidth}
    \includegraphics[width=\linewidth]{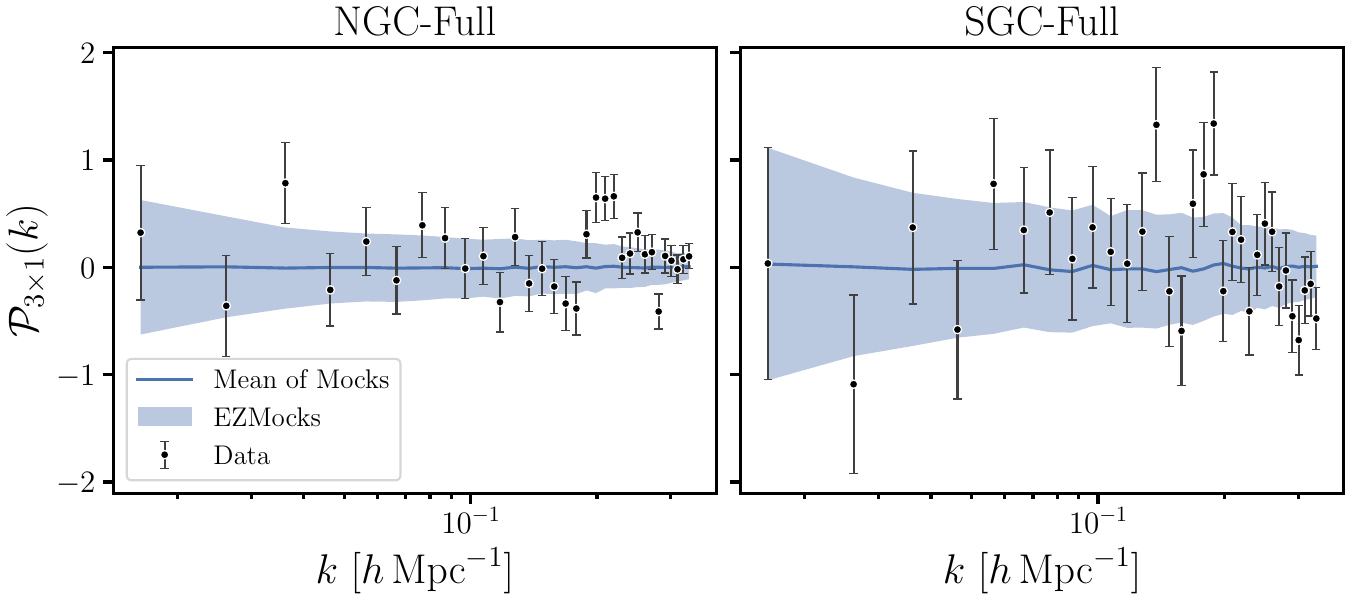}
    \end{subfigure}
    \caption{Measurements of $\mathcal{P}_{3\times 1}$ from DESI-DR1 data. The black dots are the data points, the error bars of which we estimated from the \textsc{EZmocks}. The blue solid line is the average of \textsc{EZmocks}, and the blue shaded area is their $1 \sigma$ error. }
    \label{fig:P3x1_DESI_data}
\end{figure}

\begin{figure}
    \centering
    \begin{subfigure}{0.93\textwidth}
    \includegraphics[width=\linewidth]{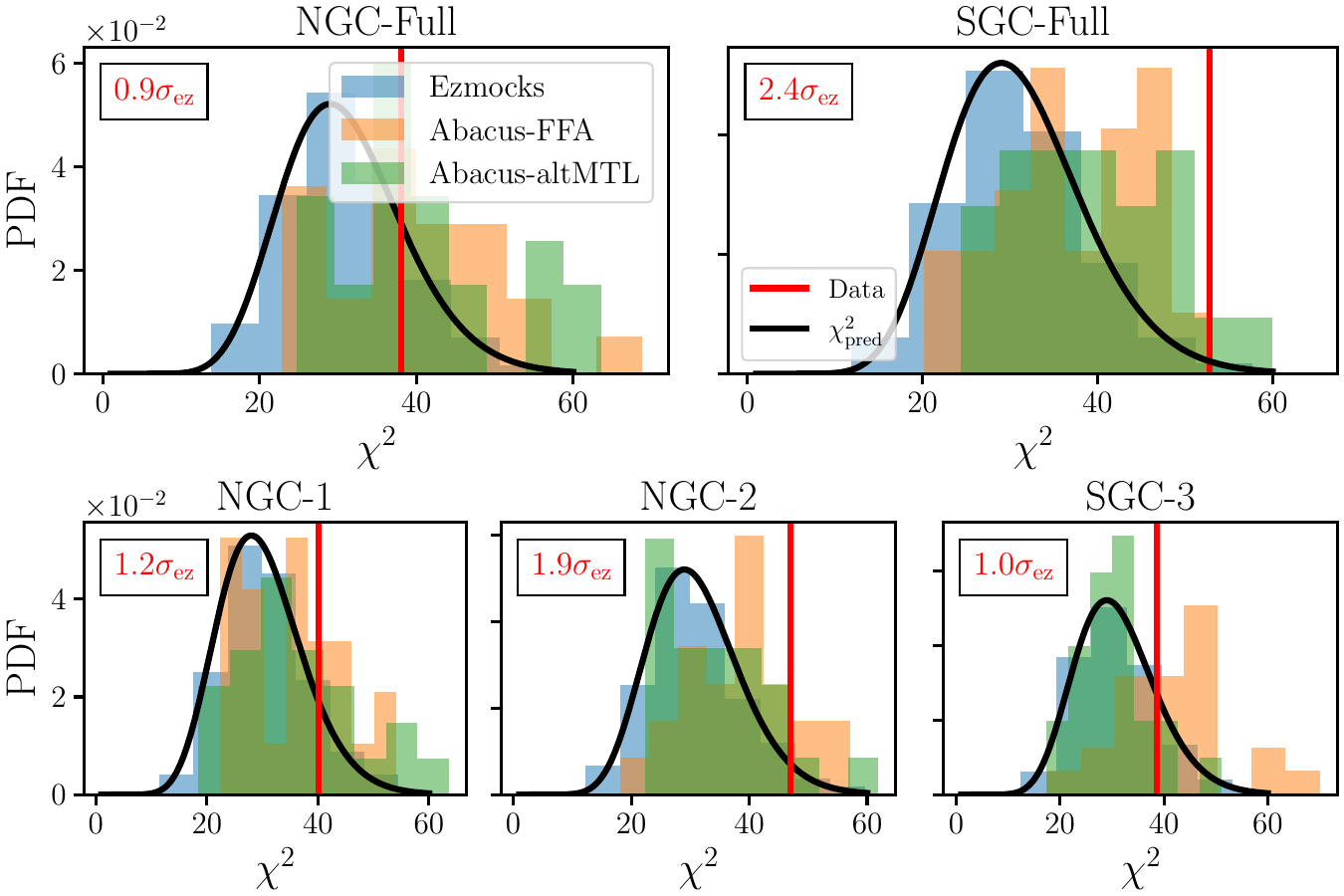}
    \caption{Chi-square for the $\mathcal{P}_{2\times 2}$ kurto spectra.}
    \label{subfig:chi2_P2x2_DESI_data}
    \end{subfigure}\vspace{0.15in}
    \begin{subfigure}{0.93\textwidth}
    \includegraphics[width=\linewidth]{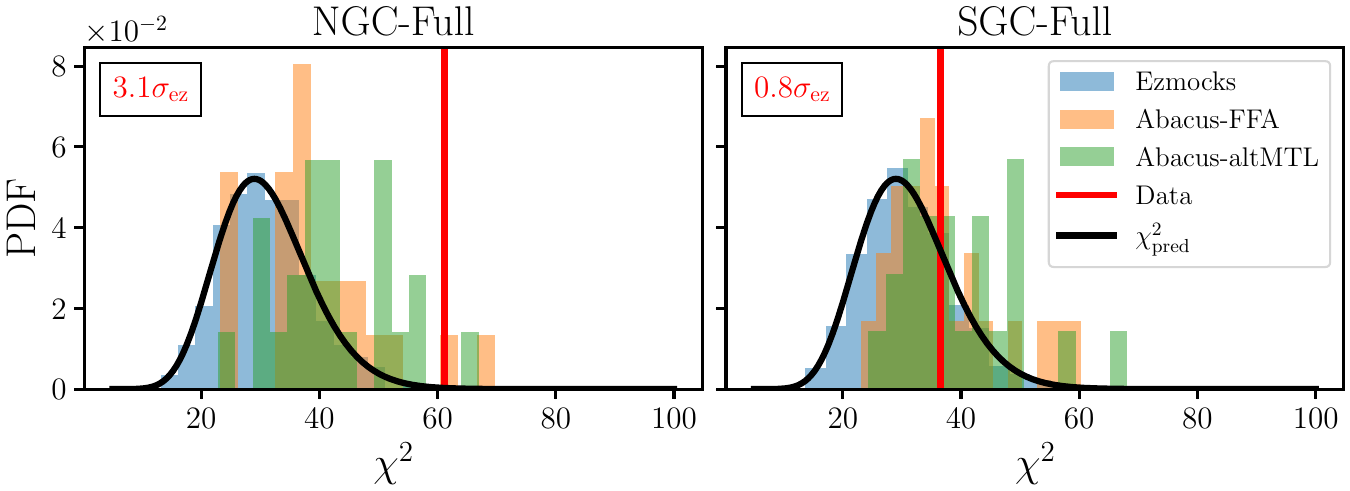}
    \caption{Chi-squared for the $\mathcal{P}_{3\times 1}$ kurto-sepctra.}
    \label{subfig:chi2_P3x1_DESI_data}
    \end{subfigure}
    \caption{The chi-square distribution of $\mathcal{P}_{2\times 2}$ and $\mathcal{P}_{3\times 1}$measurements from DESI-DR1. The covariances are estimated from \textsc{EZMOCK}. The data chi-square is presented with the red vertical line, while the distributions of \textsc{EZMOCK}, \textsc{Abacus}-FFA and \textsc{Abacus}-altMTL are shown in blue, orange and green bars, respectively. The black solid curve is the theoretical chi-squared for the given degrees of freedom.}
    \label{fig:chi2_P2x2_P3x1_DESI_data_all_patches}
\end{figure}

We show the DESI-DR1 measurements in figures~\ref{fig:P2x2_DESI_data_all_patches} and \ref{fig:P3x1_DESI_data}. The $\mathcal{P}_{2\times2}$ measurements are shown for all patches, with the full NGC and SGC footprints in the top row and the individual sub-patches in the lower row, while for $\mathcal{P}_{3\times1}$, we only show the measurements from the full patches. As in the BOSS-DR12 analysis (see figure~\ref{fig:P2x2_P3x1_BOSS_data}), the measurements fluctuate around zero, and their scatter is consistent with the $1\sigma$ mock uncertainty. Compared with the BOSS \textsc{Patchy} and \textsc{Uchuu} mocks, the scatter of the DESI \textsc{EZmock} measurements is smaller by a factor of roughly $3$--$4$. This matches the theoretical prediction based on the scaling relation of noise found in ref.~\cite{Gao:2025yqd}, where in the stochasticity-dominant regime, the variance is proportional to $\bar{n}^{-4}V^{^{-1}}$, i.e., the larger survey volume and higher mean number density reveals smaller variance. Taking NGC for example,\footnote{The volume is the cosmic volume for survey redshift range within the solid angle of survey area. The mean number density is the mean of $n(z)$ distribution shown in figure~\ref{fig:nz_comparison}.} we have two survey volumes as $V_{\rm BOSS} = 2.85\, h^{-3}\, \mathrm{Gpc}^{3}$ and $V_{\rm DESI} = 5.06\, h^{-3}\, \mathrm{Gpc}^{3}$, while the mean number densities are $\bar{n}_{\rm BOSS} = 3.1\times 10^{-4}\, h^{3}\, \mathrm{Gpc}^{-3}$ and $\bar{n}_{\rm DESI} = 4.3\times 10^{-4}\, h^{3}\, \mathrm{Gpc}^{-3}$. Additionally, we set different $k$-bins for two surveys ($\Delta k_{\rm BOSS} = 7\times 10^{-3}\, h\, \mathrm{Mpc}^{-1}$ and $\Delta k_{\rm DESI} = 1.02\times 10^{-2}\, h\, \mathrm{Mpc}^{-1}$), which contributes a factor of $\Delta k^{-1}$ to the variance. Combining all the above factors, we obtain the ratio of the standard deviation as $\sigma_{\rm BOSS}/\sigma_{\rm DESI}\sim 3.2$. The same analysis for SGC yields $\sigma_{\rm BOSS}/\sigma_{\rm DESI}\sim 4.1$. These are indeed consistent with direct comparison between error bands. However, one should be careful in comparing the amplitudes of the kurto spectra in the two surveys, since we adopt different normalisation schemes. In particular, the appropriate normalisation for the DESI-DR1 trispectrum estimator has not yet been fully established, and we therefore defer a detailed investigation in the future work. Nevertheless, the DESI measurements support the conclusion that the apparently non-zero large-scale modes seen in the BOSS-DR12 $\mathcal{P}_{2\times2}$ are not robust features shared by both surveys. From the second row of figure~\ref{fig:P2x2_DESI_data_all_patches}, we observe different levels of scatter across the DESI sub-patches, which likely reflect differences in the mean tracer number density. In particular, the NGC-2 and NGC-Full measurements show comparable scatter, and a similar behaviour is seen for the SGC-3 and SGC-Full pair. By contrast, the NGC-1 sub-patch exhibits a visibly larger scatter, consistent with its lower completeness.

We present the $\chi^2$ distributions of the $\mathcal{P}_{2\times 2}$ and $\mathcal{P}_{3\times 1}$ measurements from DESI-DR1 in figures~\ref{subfig:chi2_P2x2_DESI_data} and \ref{subfig:chi2_P3x1_DESI_data}, respectively. For $\mathcal{P}_{2\times 2}$ we also show the sub-patch measurements in the second row of figure~\ref{subfig:chi2_P2x2_DESI_data}, whereas for $\mathcal{P}_{3\times 1}$ we show only the full NGC and SGC patches. In these plots, the black curve shows the theoretical $\chi^2$ distribution. The blue, orange, and green histograms correspond to the distributions measured from the \textsc{EZMocks}, \textsc{AbacusSummit}-FFA, and \textsc{AbacusSummit}-altMTL mocks, respectively, with the covariance matrix estimated from the \textsc{EZMocks}. The red vertical line marks the data $\chi^2$ value.

The main result from these plots is that, as in BOSS-DR12, there is no detection of a PV signal. As expected, the \textsc{EZmocks} distribution (blue histogram), which is used for the covariance estimation, follows the expected theoretical $\chi^2$ with $\text{d.o.f}=31$ (for the DESI analysis $N_{\mathrm{b}}=32$, and $N_{\mathrm{s}}=1000$), providing a useful sanity check of the methodology. For $\mathcal{P}_{2\times 2}$, we find the largest deviation of data from mocks in the SGC-Full patch ($2.4\sigma$), while for the $\mathcal{P}_{3\times 1}$, we find it in the NGC-Full patch ($3.1\sigma$). We relate this observation to the non-zero data points on small scales visible in figures~\ref{fig:P2x2_DESI_data_all_patches} and \ref{fig:P3x1_DESI_data}. However, as for the large-scale outliers of BOSS-DR12, this is not an evidence for a significant detection. Hence, considering the results from all other patches, we conclude that there is no PV signal in DESI-DR1. A notable difference with respect to the BOSS-DR12 analysis is that, for DESI, the data $\chi^2$ (solid red line) is always located at the right-hand side of the mock distribution. This suggests that the mocks are a more consistent representation of the data noise for DESI than BOSS. 

\begin{figure}
    \centering
    \begin{subfigure}{0.93\textwidth}
    \includegraphics[width=\linewidth]{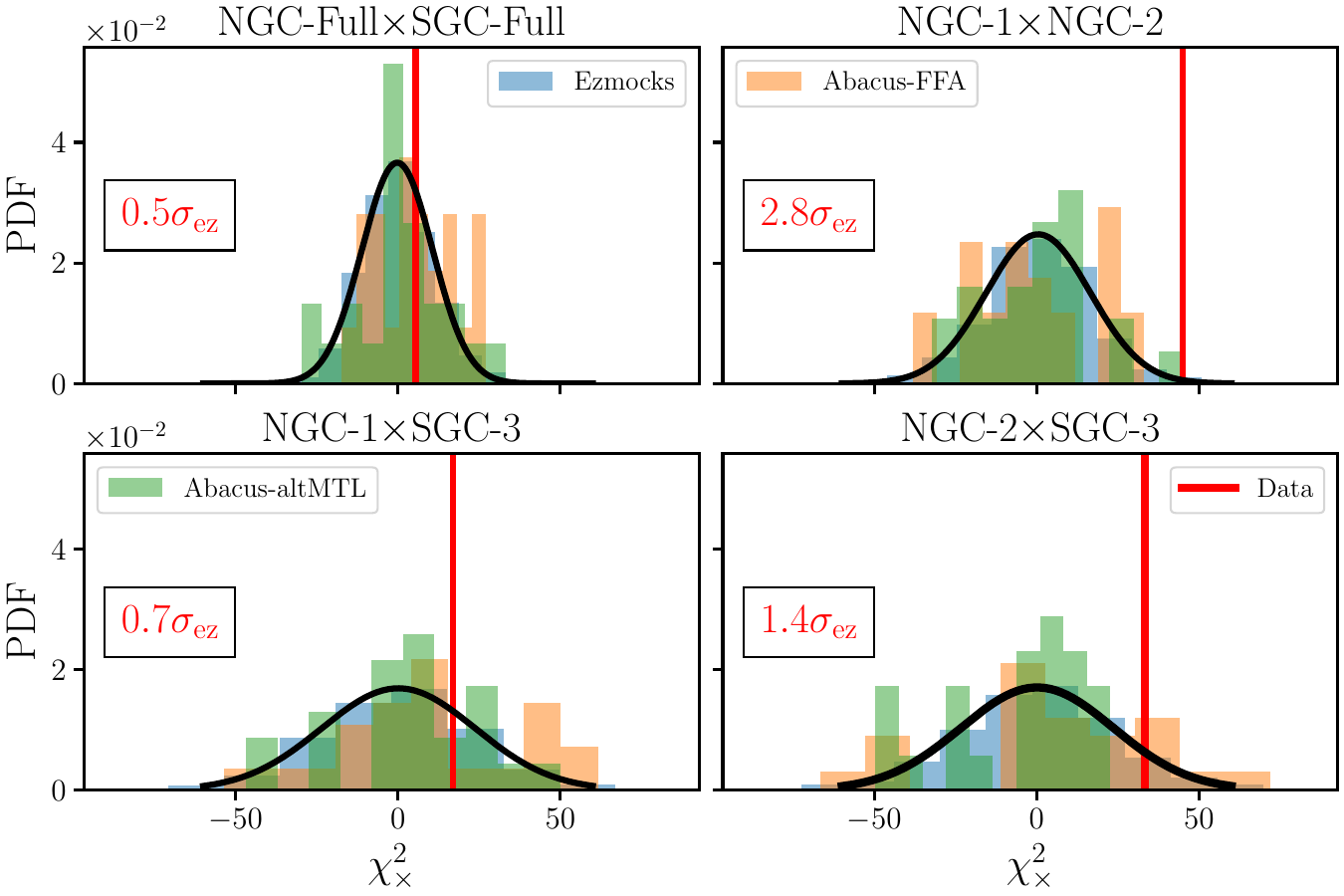}
    \caption{NGC-Full Covariance}
    \label{subfig:chi_cross_P2x2_DESI_NGC}
    \end{subfigure}
    \begin{subfigure}{0.93\textwidth}
    \includegraphics[width=\linewidth]{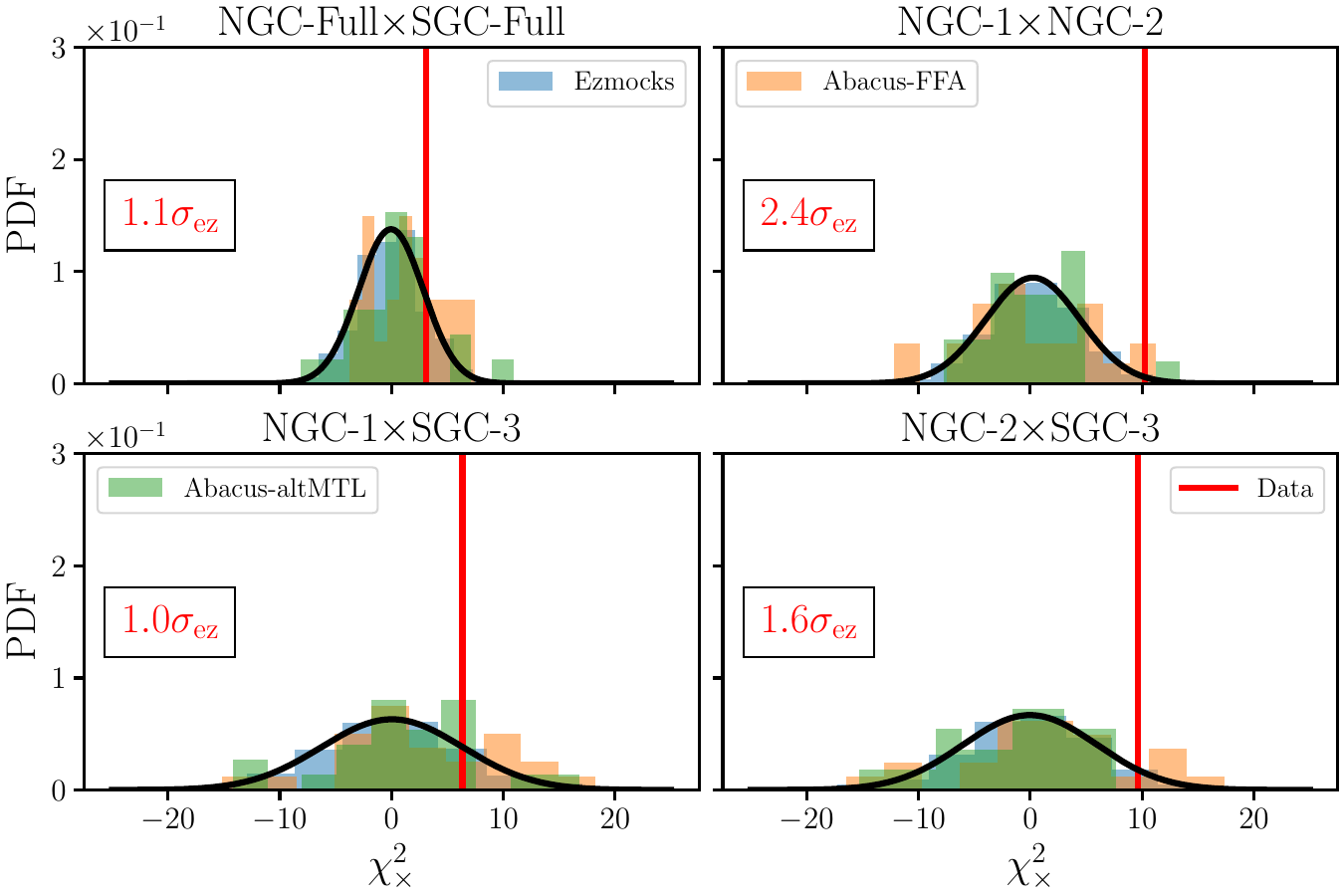}
    \caption{SGC-Full Covariance}
    \label{subfig:chi_cross_P2x2_DESI_SGC}
    \end{subfigure}
    \caption{The $\chi^2_\times$ statistics of $\mathcal{P}_{2\times 2}$ measurements from DESI-DR1. We used different covariances in the two sub-figures (see captions). The colour coding is the same as figure~\ref{fig:chi2_P2x2_P3x1_DESI_data_all_patches} except for the black curve, which is a Gaussian distribution centred on $0$ and standard deviation computed from the \textsc{EZmocks} distribution.}
    \label{fig:chi_cross_P2x2_DESI_data_all_patches}
\end{figure}

\begin{figure}
    \centering
    \includegraphics[width=0.6\linewidth]{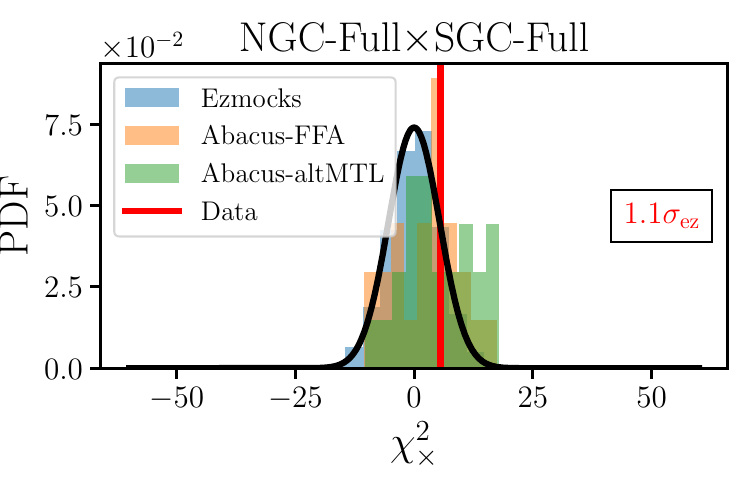}
    \caption{The $\chi^2_\times $ statistics of $\mathcal{P}_{3\times 1}$ measurements from DESI using covariance estimated from NGC-Full. The colour coding is the same as figure~\ref{fig:chi_cross_P2x2_DESI_data_all_patches}.}
    \label{fig:chi_cross_P3x1_DESI_data_all_patches}
\end{figure}
For the $\chi^2$ statistic, both \textsc{AbacusSummit} mock sets are shifted toward larger values than the \textsc{EZMocks} distribution, indicating that the kurto spectra measured from the \textsc{AbacusSummit} mocks are more scattered when assessed with the \textsc{EZMocks} covariance. The difference between the FFA and altMTL fibre-assignment schemes is small, suggesting that the kurto-spectrum estimator is not strongly sensitive to fibre assignment. This is consistent with the construction of the composite fields, where fluctuations on scales below $7\,h^{-1}\,\mathrm{Mpc}$ are smoothed. Given the limited number of \textsc{AbacusSummit} realisations, however, this comparison should be viewed as indicative rather than definitive. The data $\chi^2$ values are generally more compatible with the \textsc{AbacusSummit} distributions than with the \textsc{EZMocks} distribution. Since the \textsc{AbacusSummit} mocks are based on full $N$-body simulations, whereas the \textsc{EZMocks} rely on an approximate method, this trend is qualitatively reasonable if no PV signal is present in the data. Part of the broader \textsc{AbacusSummit} distribution could also arise from the box replication used to fill the DESI volume~\cite{Hou:2025cey}, although the present level of agreement with the data suggests that this effect is unlikely to dominate.

Figures~\ref{fig:chi_cross_P2x2_DESI_data_all_patches} and \ref{fig:chi_cross_P3x1_DESI_data_all_patches} show the corresponding $\chi^2_{\times}$ statistics. These lead to the same overall conclusion: for all patch combinations, the data are consistent with the null distribution derived from the \textsc{EZMocks}, with no deviation exceeding $3\sigma$. The largest deviation is found for the NGC-1$\times$NGC-2 cross-correlation, reaching about $2.8\sigma$ when the NGC-Full covariance is adopted and about $2.4\sigma$ when the SGC-Full covariance is used. A plausible explanation is that NGC-1 and NGC-2 lie in the same sky region and may therefore share residual observational systematics, which could enhance their apparent cross-correlation.

In subfigures~\ref{subfig:chi_cross_P2x2_DESI_NGC} and \ref{subfig:chi_cross_P2x2_DESI_SGC}, we compute $\chi^2_{\times}$ using the NGC-Full and SGC-Full covariances, respectively. The comparison shows that the SGC-Full covariance is larger, as expected from its smaller effective survey volume and slightly lower number density. Consequently, the absolute value of $\chi^2_{\times}$ is generally reduced when the SGC-Full covariance is used, so the data point moves closer to zero. At the same time, the mock null distribution becomes more narrowly concentrated around zero, so the significance does not necessarily decrease even though the raw $\chi^2_{\times}$ value becomes smaller. The only case in which the significance decreases when switching to the SGC-Full covariance is NGC-1$\times$NGC-2, again suggesting that its relatively large deviation is tied to effects specific to the NGC region and is better captured by the NGC-based covariance.

For $\chi^2_{\times}$, the three mock families behave similarly and remain centered around the null expectation, as expected in the absence of a PV signal. This suggests that the cross-patch statistic is not strongly affected by the differences among the mock families considered here.

\begin{figure}
    \centering
    \includegraphics[width=1.\linewidth]{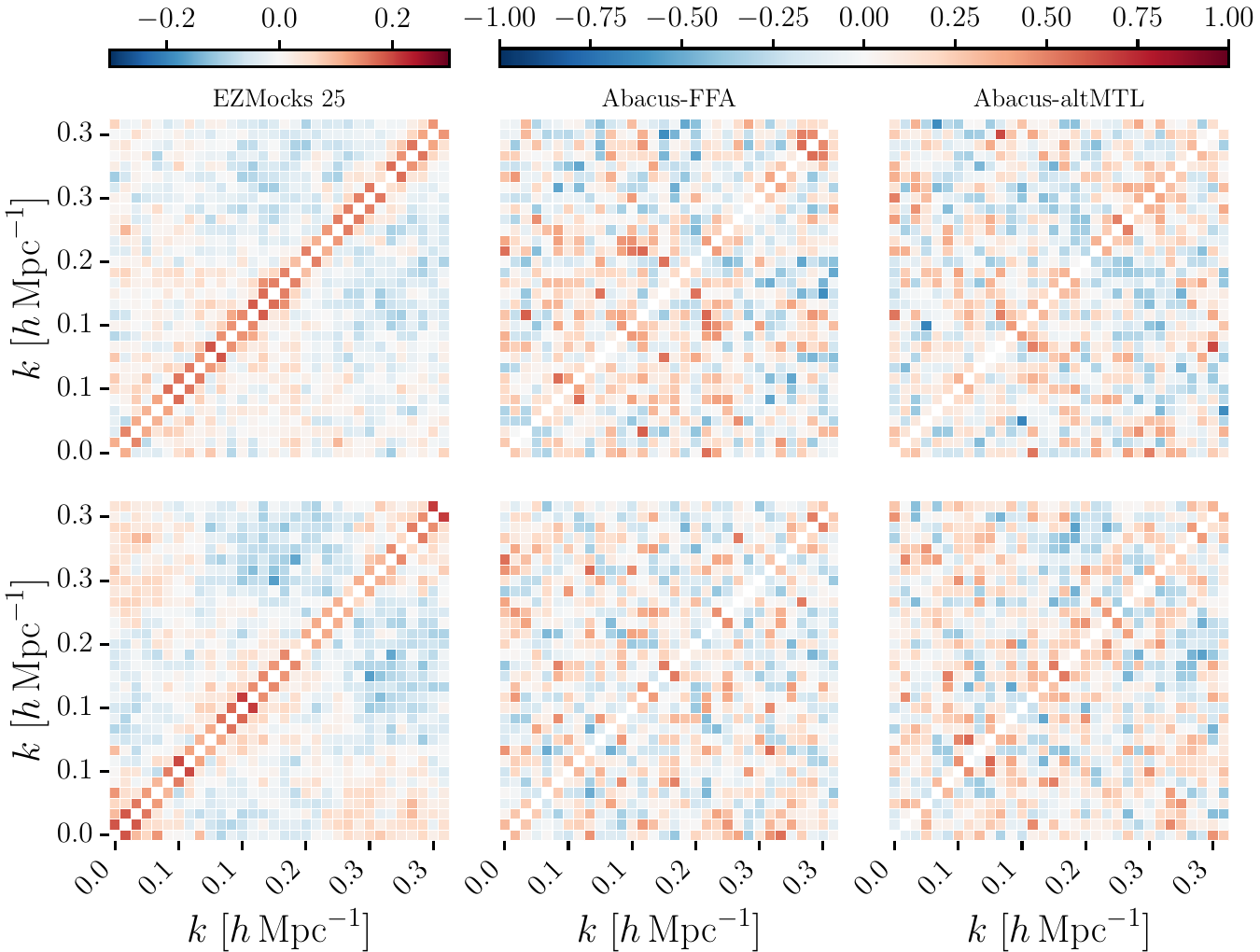}
    \caption{Correlation matrices for DESI DR1 NGC. The top row shows the matrices for $\mathcal{P}_{2\times 2}(k)$, and the bottom one for $\mathcal{P}_{3\times 1}(k)$. In the left, middle, and right columns we present the \textsc{EZMocks}, the \textsc{AbacusSummit}-FFA and the  \textsc{AbacusSummit}-altMTL correlation matrix, respectively. We note that the colorbar ranges of the first column is different from that of the last two columns, given the numerical covariance of \textsc{EZMocks} has very small off-diagonal elements. We note that the diagonal elements are masked out.}
    \label{fig:correlation-matrix-desi}
\end{figure}

We present the correlation matrices for the three mocks in figure~\ref{fig:correlation-matrix-desi}. The covariance for \textsc{AbacusSummit} shown here is estimated numerically from the $25$ mocks, which are not used for $\chi^2$ analyses. Since the number of mocks is smaller than the dimensionality of the data vector, the covariance is noisy. The \textsc{EZMocks} shows small off-diagonal feature, as found also in BOSS case~\ref{fig:correlation-matrix}. On the contrary, the off-diagonal amplitudes of \textsc{AbacusSummit} are not negligible, which originates from the noisy estimation of covariance resulting from limited mock number\footnote{We have tested the covariance estimation using $25$ \textsc{EZMocks}, yielding the same large off-diagonal amplitude as \textsc{AbacusSummit} mocks.}.

\section{Conclusion}\label{sec:conclusion}

We have presented the first measurement of parity-odd-sensitive kurto spectra~\cite{Jamieson:2024mau,Gao:2025yqd} on spectroscopic galaxy survey data. We considered two estimators, $\mathcal{P}_{2\times2}$ defined in eq.~\eqref{eq:P_2x2} and $\mathcal{P}_{3\times1}$ defined in eq.~\eqref{eq:P_3x1}, which are power-spectrum-like composite-field estimators constructed to have a non-zero expectation value in the presence of a parity-violating signal, and applied them to the BOSS-DR12 and DESI-DR1 data sets. For each survey, we measured the same observables on the corresponding mock catalogues and estimated their covariance matrices numerically. A central advantage of the kurto-spectrum approach is its much lower dimensionality relative to the full parity-odd 4PCF, which makes numerical covariance estimation both feasible and robust.

Using null-hypothesis tests based on the $\chi^2$ statistic defined in eq.~\eqref{eq:chi_square}, we find no evidence for a parity-violating signal in either BOSS-DR12 or DESI-DR1. This conclusion is stable across the different mock suites considered in this work. We also analysed the cross-patch statistic $\chi^2_{\times}$ introduced in ref.~\cite{Krolewski:2024JCAP}, which is designed to isolate a signal shared between patches while reducing sensitivity to data--mock mismatch. The measured $\chi^2_{\times}$ values are likewise consistent with the null expectation in both surveys, further supporting the absence of a statistically significant PV signal.

We also investigated the impact of the adopted mock catalogues on the inferred significance. For both BOSS-DR12 and DESI-DR1, the $\chi^2$ distributions of the higher-fidelity simulations (\textsc{Uchuu} for BOSS and \textsc{AbacusSummit} for DESI) are generally more consistent with the data than those of the approximate mock suites. For DESI-DR1, we further find that the kurto spectra are only weakly sensitive to the choice of fibre-assignment scheme, consistent with the smoothing applied to suppress small-scale fluctuations. The effect of box replication in the \textsc{AbacusSummit} mocks also appears to be subdominant at the present level of precision.

Comparing the data vectors and mock error bands of $\mathcal{P}_{2\times2}$ and $\mathcal{P}_{3\times1}$, we find that the scatter in DESI-DR1 is about a factor of four smaller than in BOSS-DR12. This indicates a substantial improvement in statistical sensitivity, consistent with the higher number density and larger effective volume of DESI. A null result in both surveys, therefore, implies that any cosmological parity-violating kurto-spectrum signal must lie below the current DESI-DR1 noise level. Future DESI data releases, with larger volume and higher completeness, should significantly improve these bounds.

By studying the correlation matrices, we find that the covariance of the kurto spectra is close to diagonal in both surveys, with only weak off-diagonal correlations. This is consistent with a noise-dominated regime and is another practical advantage of the kurto-spectrum estimator. For BOSS-DR12, we also find that the \textsc{Uchuu} mocks have systematically smaller variance than \textsc{Patchy}, opposite to the trend reported for the parity-odd 4PCF in ref.~\cite{Philcox:2024mmz}. This further illustrates that the relative behaviour of different mock suites is statistic-dependent and need not be the same for configuration-space and Fourier-space observables.

Overall, this work shows that parity-odd-sensitive kurto spectra can be measured robustly on real survey data and provide a valuable complement to parity tests based on the 4PCF. They offer both an independent cross-check of existing 4PCF analyses and a numerically tractable route for future searches. In particular, the lower dimensionality of the estimator makes direct numerical covariance estimation feasible, which is a major practical advantage over previous 4PCF approaches. Moreover, another advantage of the kurto-spectra formalism is that it compresses the trispectrum in a more physically motivated way, i.e. tuning the operators based on theoretical trispectrum templates. 

There are several clear directions for improving and extending the present analysis. On the observational side, DESI-DR1 is still incomplete, and future data releases with larger volume and improved completeness will provide a more stringent test of parity violation. On the modelling side, the limited number of high-fidelity DESI mocks currently restricts a more detailed investigation of systematic effects and mock dependence. In addition, residual data--mock mismatch at the level of the trispectrum may still affect the inferred significance, even though the kurto spectra appear more robust than the full 4PCF. The estimator itself can also be improved. The normalisation of the trispectrum estimator warrants further study, particularly because the standard survey weights are not optimised for trispectrum measurements. More generally, the definition of the kurto spectra could be refined through improved weighting schemes, for example inverse-variance or luminosity-dependent weights, with the aim of suppressing the parity-even stochastic contribution to the noise~\cite{Fang:2023ypv,Gao:2025yqd}. A further important direction is to extend the kurto-spectrum framework beyond the generic null test considered here and build a bank of separable parity-odd templates, for example for massive-spin exchange, axion--gauge scenarios, and helical primordial magnetic fields. This would make it possible to identify which classes of models are best constrained by current and future data, and to clarify how kurto-spectrum analyses complement searches based on the full trispectrum. With larger DESI samples, improved mock suites, and further refinement of the estimator, kurto-spectrum measurements should provide increasingly stringent tests of parity violation in large-scale structure.

\acknowledgments
We thank Jiamin Hou and Oliver Philcox for discussions related to trispectrum analysis of BOSS and DESI data and their feedback on the early draft of this manuscript. 
We also thank Davide Bianchi and Seshadri Nadathur for discussions about DESI data. The work of LAPTh group is supported by the Agence Nationale de la Recherche (ANR) under grant No. ANR-23-CPJ1-0160-01. ZV acknowledges the support of the Croatian Science Foundation (HRZZ) under the project grant IP-2025-02-1338. All measurements and analyses were performed using the computing and storage resources of GENCI at IDRIS, thanks to allocations 2025-AD010416295 and 2025-AD010417047 on the CSL partition of the Jean Zay supercomputer. 

\appendix 

\section{Proofs on signal sensitive cross statistics}\label{append:chi-cross}
\begin{figure}
    \centering
    \includegraphics[width=0.92\linewidth]{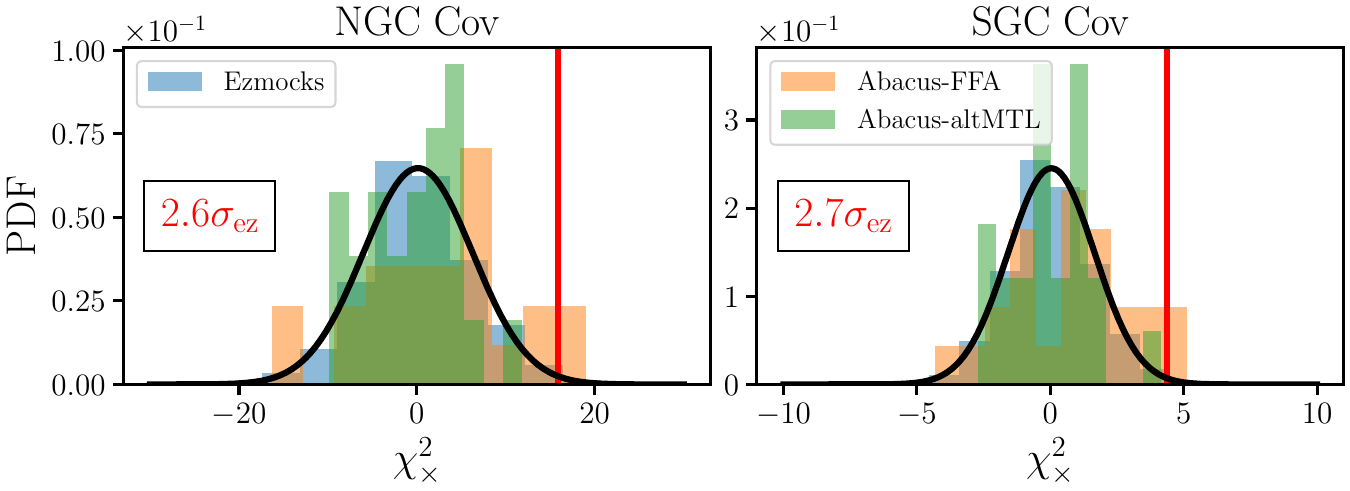}
    \caption{The $\chi_{\times}^2$ calculated by combining NGC-1, NGC-2 and SGC-3. The left and right panels are using the covariance matrix from the estimated \textsc{EZMocks} NGC-full and SGC-Full, respectively. The legends are the same as figure~\ref{fig:chi_cross_P2x2_DESI_data_all_patches}.}
    \label{fig:chi_cross_sum}
\end{figure}

We present the result of $\chi^2_{\times}$ (eq.~\ref{eq:chi_cross}) by combining all the DESI sub-patches in figure~\ref{fig:chi_cross_sum}. The deviation from zero of the data $\chi_\times^2$ is similar to that of NGC-1 $\times $ NGC-2. We argue that this NGC cross patch is the main driver of this large discrepancy. 

We revisit the proof of why $\chi_\times^2$ is insensitive to the data-mock mismatch.\footnote{A recapitulation on the meaning of data-mock mismatch: the mocks are not calibrated to match higher-order statistics in data. In the null hypothesis test, the difference in $\chi^2$ can originate from non-zero $\mathrm{Tr}(\mathbf{C}_{\mathrm{data}} \mathbf{C}_{\mathrm{mock}}^{-1} - \mathbf{1})$~\cite{Krolewski:2024JCAP}.} Following ref.~\cite{Krolewski:2024JCAP}, we first assume that the data vector is an unbiased estimator for the cosmic PV signature. For patch $\mu$, we have
\begin{equation}
    \langle \xi^{\mu}_a \rangle_{\mathrm{data}} = \bar{\xi}_a, \hspace{1cm} \langle \xi^{\mu}_a \rangle_{\mathrm{mock}} = 0,
\end{equation}
where we used the fact that mocks do not have PV signature. Considering the cross-patch correlations, we have
\begin{equation}
    \langle \xi^\mu_a \xi^\nu_b \rangle_{\mathrm{data}} = \bar{\xi}_a \bar{\xi}_b + \mathbf{C}^{\mu \nu}_{ab,\,\mathrm{data}}, \hspace{1cm} \langle \xi^\mu_a \xi^\nu_b \rangle_{\mathrm{mock}} = \mathbf{C}^{\mu \nu}_{ab,\,\mathrm{mock}},
\end{equation}
where $\mathbf{C}^{\mu\nu}$ is the cross covariance between two patches. In ref.~\cite{Krolewski:2024JCAP,Hou:2025cey}, this cross covariance is assumed to be zero if $\mu \neq \nu$. Without this assumption, the expectation value of $\chi_\times^2$ for data and mocks can be generically written as
\begin{align}
    &\langle \chi_\times^2 \rangle_{\mathrm{data}} = \frac{1}{N_p(N_p-1)}\sum_{\mu < \nu} \hat{\mathbf{C}}^{-1}_{ab} \langle \xi^\mu_a \xi^\nu_b \rangle_{\mathrm{data}} = \bar{\xi}_a\hat{\mathbf{C}}^{-1}_{ab}\bar{\xi}_b + \frac{1}{N_p(N_p-1)} \hat{\mathbf{C}}^{-1}_{ab} \sum_{\mu < \nu} \mathbf{C}^{\mu \nu}_{ab,\,\mathrm{data}}, \\
    &\langle \chi_\times^2 \rangle_{\mathrm{mock}} = \frac{1}{N_p(N_p-1)} \hat{\mathbf{C}}^{-1}_{ab} \sum_{\mu < \nu} \mathbf{C}^{\mu \nu}_{ab,\,\mathrm{mock}}.
\end{align}

From figures~\ref{fig:chi_cross_P2x2_DESI_data_all_patches} and~\ref{fig:chi_cross_P3x1_DESI_data_all_patches}, we confirm that $\langle \chi_\times^2 \rangle_{\mathrm{mock}}$ has a zero mean with a certain spread, demonstrating that the cross-patch covariance for the mock is indeed zero. However, the data $\chi_\times^2$ can still follow a different distribution without the presence of a PV signature, for the last two aspects discussed in section~\ref{sec:chicross}. 

\section{Effect of $k$-cut on null hypothesis}\label{append:k_cut}

\begin{figure}
    \centering
    \begin{subfigure}{0.93\textwidth}
    \includegraphics[width=\linewidth]{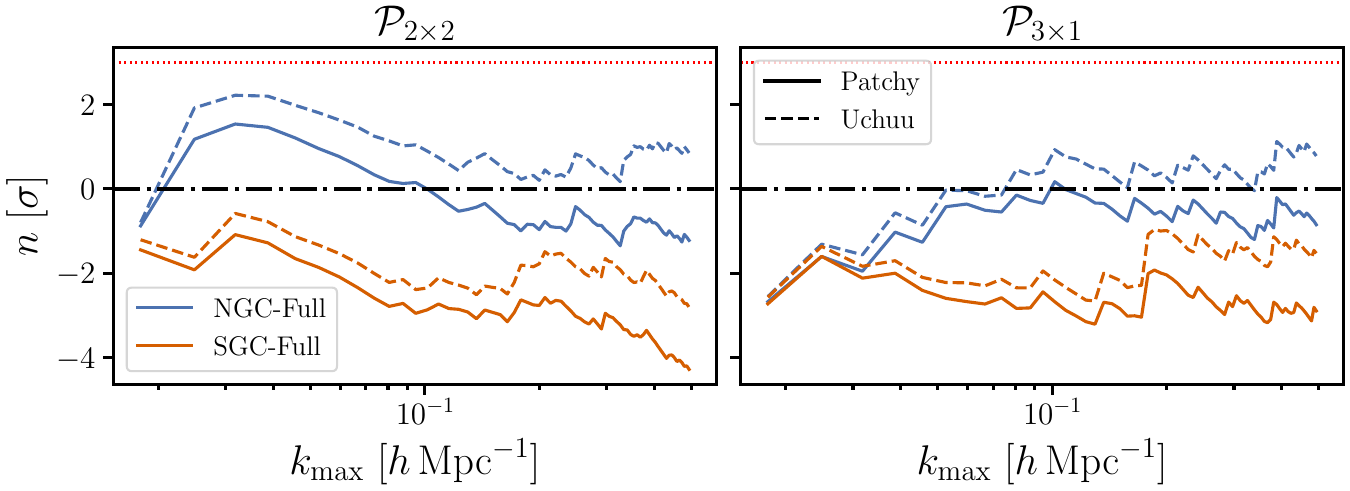}
    \caption{BOSS}
    \label{subfig:z_score_k_cut_BOSS}
    \end{subfigure}
    \begin{subfigure}{0.93\textwidth}
    \includegraphics[width=\linewidth]{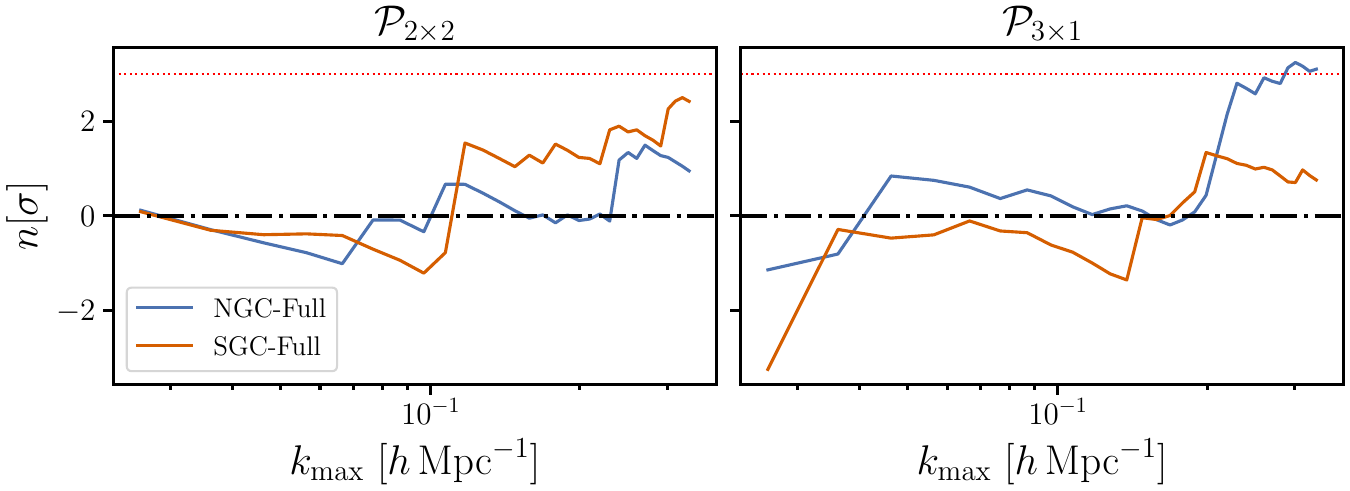}
    \caption{DESI}
    \label{subfig:z_score_k_cut_DESI}
    \end{subfigure}
    \caption{The significance of data $\chi^2$ compared with mock $\chi^2$-distribution. The two subfigures separately show the results from BOSS and DESI. In all panels, the blue and orange lines correspond to NGC-Full and SGC-Full analyses, while the left and right panels correspond to $\mathcal{P}_{2\times 2}$ and $\mathcal{P}_{3\times 1}$. For BOSS, the solid and dashed lines represent using \textsc{Patchy} and \textsc{Uchuu} mocks. The red dotted line is the $3\sigma$ threshold.}
    \label{fig:z_score_k_cut_BOSS_DESI}
\end{figure}

In figure~\ref{fig:z_score_k_cut_BOSS_DESI}, we demonstrate the results for the null hypothesis test applying different $k$-cuts. The y-axis shows the signed significance, in units of $\sigma$, of the data $\chi^2$ relative to the mock $\chi^2$ distribution. For each value of $k_{\rm max}$, the $\chi^2$ for data and mocks is calculated using the data vectors truncated respectively. The last points in these plots are the sigma-levels quoted in the main text. 

For BOSS (figure~\ref{subfig:z_score_k_cut_BOSS}), the significance is generally negative, except in two cases: for the NGC measured with \textsc{Uchuu} mocks in both kurto spectra, and for $\mathcal{P}_{2\times2}$ in the NGC with \textsc{Patchy} when only large-scale modes with $k\lesssim 0.1\,h\,\mathrm{Mpc}^{-1}$ are included. The first case is consistent with our earlier finding that the \textsc{Uchuu} mocks provide a more accurate description of the data fluctuations than the \textsc{Patchy} mocks. The second reflects the visibly non-zero large-scale modes in $\mathcal{P}_{2\times2}$ (see figure~\ref{fig:P2x2_P3x1_BOSS_data}). More generally, the tendency for the significance to remain negative or decrease as smaller scales are included suggests that the mocks tend to overestimate the small-scale variance of the data. Nevertheless, none of the BOSS curves exceeds the $3\sigma$ threshold required for a significant detection.

For DESI (figure~\ref{subfig:z_score_k_cut_DESI}), the significance curves show a more irregular dependence on $k_{\max}$, with several abrupt changes as additional modes are included. In particular, the final points of $\mathcal{P}_{3\times1}$ approach the $3\sigma$ level. However, the sharp jumps around $k\sim 0.2\,h\,\mathrm{Mpc}^{-1}$ indicate that the result is sensitive to a small number of modes and is therefore not stable under changes of the scale cut. For this reason, we do not interpret these features as evidence for a significant non-zero signal. This is consistent with the conclusion from the $\chi^2_{\times}$ analysis that the DESI measurements remain compatible with the null hypothesis.

\bibliographystyle{JHEP}
\bibliography{biblio.bib}
\end{document}